\documentclass[a4paper,11pt]{article}
\pdfoutput=1
\usepackage{jcappub} 
\usepackage[T1]{fontenc} 
\usepackage{subcaption}
 
\definecolor{mypink1}{rgb}{0.858, 0.188, 0.478}

\title{Implications of the {\it Fermi}-LAT Pass 8 Galactic Center Excess on Supersymmetric Dark Matter}
\author[a]{Abraham Achterberg}
\author[a]{Melissa van Beekveld}
\author[a,b]{Sascha Caron}
\author[c]{Germ\'an A. G\'omez-Vargas}
\author[a]{Luc Hendriks}
\author[d]{Roberto Ruiz de Austri}
\affiliation[a]{Institute for Mathematics, Astrophysics and Particle Physics, Faculty of Science, Mailbox 79,
Radboud University Nijmegen, P.O. Box 9010, NL-6500 GL Nijmegen, The Netherlands}
\affiliation[b]{Nikhef, Science Park, Amsterdam, The Netherlands}
\affiliation[c]{Instituto de Astrof\'isica, Pontificia Universidad Cat\'olica de Chile, Avenida Vicu\~na Mackenna 4860, Santiago, Chile}
\affiliation[d]{Instituto de F\'isica Corpuscular, IFIC-UV/CSIC, Valencia, Spain}

\emailAdd{a.achterberg@astro.ru.nl}
\emailAdd{mcbeekveld@gmail.com}
\emailAdd{scaron@cern.ch}
\emailAdd{ggomezv@uc.cl}
\emailAdd{luc@luchendriks.com}
\emailAdd{rruiz@ific.uv.es}

\abstract{The {\it Fermi} Collaboration has recently updated their analysis of gamma rays from the center of the Galaxy. They reconfirm the presence of an unexplained emission feature which is most prominent in the region of $1-10$ GeV, known as the Galactic Center GeV excess (GCE). Although the GCE is now firmly detected, an interpretation of this emission as a signal of self-annihilating dark matter (DM) particles is not unambiguously possible due to systematic effects in the gamma-ray modeling estimated in the Galactic Plane. In this paper we build a covariance matrix, collecting different systematic uncertainties investigated in the {\it Fermi} Collaboration's paper that affect the GCE spectrum.
We show that models where part of the GCE is due to annihilating DM is still 
consistent with the new data. We also re-evaluate the parameter space regions of the minimal supersymmetric Standard Model (MSSM) that can contribute dominantly to the GCE
via neutralino DM annihilation. 
All recent constraints from DM direct detection experiments such as PICO, LUX, PandaX and Xenon1T, limits on the annihilation cross section from dwarf spheroidal galaxies and the Large Hadron Collider limits are considered in this analysis. Due to a slight shift in the energy spectrum of the GC excess with respect to the previous {\it Fermi} analysis, and the recent limits from direct detection experiments, we find a slightly shifted parameter region of the MSSM, compared to our previous analysis, that is consistent with the GCE. Neutralinos with a mass between $85-220$ GeV can describe the excess via annihilation into a pair of $W$-bosons or top quarks. Remarkably, there are models with low fine-tuning among the regions that we have found. The complete set of solutions will be probed by upcoming direct detection experiments and with dedicated searches in the upcoming data of the Large Hadron Collider.
}

\begin{document} 
\maketitle
\flushbottom 

\section{Introduction}
\label{sec:introduction}

There is overwhelming evidence that the matter content of the Universe mostly consists of dark matter (DM). What still remains unknown is its nature, that is, whether it is a fundamental particle and how it interacts with the Standard Model (SM) of fundamental interactions. There are many extensions of the SM that predict particles that could account for the DM we observe. Among them, the so-called weakly interactive massive particles (WIMPs) are the more popular ones.  
WIMPs naturally achieve the required relic density through self-annihilation in the early Universe \cite{Bertone:2004pz}. Precisely this self-annihilation mechanism would allow indirect DM detection in the present Universe by observing the stable annihilation products, such as gamma rays.  In our galaxy this predicted gamma radiation could be detected by the Large Area telescope (LAT), on board the {\it Fermi} Gamma-ray Space Telescope \cite{2009ApJ...697.1071A}.   

The Galactic Center Excess (GCE) is a feature in the gamma-ray data collected from the Inner Galaxy with the {\it Fermi}-LAT. The dominant contribution of the total emission detected can be explained using Interstellar Emission Models (IEMs) tuned with Galactic plane data and point source catalogs. The GCE is a sub-dominant component ($\sim$10\%) of the observed flux.  The spectral energy distribution of the GCE peaks at about 3 GeV \cite{Goodenough:2009gk, Vitale:2009hr, Hooper:2010mq, Gordon:2013vta, Hooper:2011ti, Daylan:2014rsa, 2011PhLB..705..165B, Calore:2014xka, Abazajian:2014fta,Zhou:2014lva}. Among different IEMs and source lists, the GCE is well described spatially with a generalized Navarro, Frenk and White (NFW) dark matter density profile \cite{NFW:1997,Navarro:1995iw} with reduced inner slope (index 1.25). 

Two {\it Fermi}-LAT Collaboration papers have confirmed the GCE \cite{TheFermi-LAT:2015kwa, 1704.03910}. In particular, the latest work presents an updated status of the GCE using the reprocessed Pass 8 event data collected in about 6.5 years of observations\footnote{The data of the analysis in \cite{1704.03910}  is publicly released here https:/www-glast.stanford.edu/pub\_data/1220/} \cite{1704.03910}. A large set of systematic sources of uncertainty in the extraction of the GCE properties (spectral shape, magnitude and morphology) were investigated in ref. \cite{1704.03910}. This results in an excess spectrum that is most prominent in the photon energy range of 1-10 GeV and has a smaller high energy tail starting at photon energies of 10 GeV. The spatial morphology of the GCE in these two energy ranges seem to be different from one another, the first energy range being compatible with a DM signal and at energies larger than about 10 GeV with an X-shaped morphology is instead observed~\cite{1704.03910}. One concludes that the GCE is detected but that, given the magnitude of systematic uncertainties, a firm interpretation of the excess as a signal from DM annihilation is not possible, but also not excluded.

A possible interpretation of the GCE comes from the fact that 205 pulsars have been identified in the gamma-ray band\footnote{See https://confluence.slac.stanford.edu/x/5Jl6Bg for the list of detected $\gamma$-ray point pulsars.}.  A population of pulsars is expected in the Galactic bulge \cite{Macquart:2015jfa}.  Based on these assumptions many previous studies have investigated the possibility that a population of  unresolved gamma-ray pulsars could be the reason of the GCE emission \cite{McCann:2014dea,Mirabal:2013rba,OLeary:2015qpx,Yuan:2014rca,Petrovic:2014xra,Cholis:2014lta}. For instance, ref. \cite{Cholis:2014lta} concludes that about 60 Galactic bulge pulsars should have been already seen by the {\it Fermi}-LAT, but have not been identified as pulsars. Furthermore, using novel statistical methods, the authors of ref. \cite{2016PhRvL.116e1103L} and ref. \cite{2016PhRvL.116e1102B} have claimed evidence for the existence of an unresolved population of gamma-ray sources in the inner $20\deg$ of the Galaxy, with a spatial distribution and collective flux compatible with the GCE.  

Recently the {\it Fermi}-LAT Collaboration investigated the pulsar interpretation of the GCE \cite{1704.03910}. Performing a new point source search in 7.5 years of Pass 8 {\it Fermi}-LAT data in a $40^{\text{o}} \times  40^{\text{o}}$ box around the GC, they  claim to confirm the findings of ref. \cite{2016PhRvL.116e1103L} and ref. \cite{2016PhRvL.116e1102B}. In this analysis they detect 400 sources, with 66 of them being gamma-ray pulsar candidates. They also find that these sources are more likely to be the brighter members of a larger underlying population of pulsars in the Galactic bulge rather than Galactic bar pulsars. They find that the collective emission of gamma-ray pulsars in the bulge population is compatible with the GCE properties. However, some arguments \cite{Hooper:2015jlu,Hooper:2016rap,Haggard:2017lyq} have been raised against this explanation, pointing out that a gamma-ray millisecond pulsar (MSP) population in the Galactic bulge with similar properties of already confirmed populations in globular clusters\footnote{For instance in ref. \cite{Haggard:2017lyq}, assuming that the ratio between low mass x-ray binaries to MSPs in globular clusters is the same as in the bulge, it is shown that the MSP contribution to the GCE emission is limited to 4\% to 23\%.} and the local Galactic disc\footnote{The authors of \cite{Hooper:2016rap} argue that if the luminosity function of local Galactic disc MSPs is assumed for the bulge population, the {\it Fermi}-LAT should have detected many of its bright members in the GC region, but none have been identified yet.}, is not able to reproduce the whole GCE emission. The weak point of those arguments is the assumption that a MSP population in the Galactic bulge shares similarities with populations in different environments and with a diverse origin, such as in globular clusters and the local Galactic disc \cite{Ploeg:2017vai}. Therefore the debate is not yet closed, leaving the possibility that the entire GCE or a fraction of the GCE is due to DM (WIMP) annihilation. Therefore, new methods are needed to analyze $\gamma$-ray data (see~\cite{Caron:2017udl}) to detect the pulsar population at other wavelengths and to determine the true nature of the excess.
There have been many attempts in the literature to fit the GCE through WIMPs, mainly using simplified models. These simplified models do not cover the full phenomenology of more complete and more complex models,  e.g. the only possible signature at the LHC could come from the production of a particle which is not present in the simplified model.
Therefore Supersymmetry (SUSY) in its minimal phenomenological realization, the so called phenomenological Minimal Supersymmetric Standard Model (pMSSM) \cite{Djouadi:1998di}, has also been investigated in ref. \cite{Caron:2015wda,Bertone:2015tza,Butter:2016tjc,Freese:2015ysa,Agrawal:2014oha,Cerdeno:2015ega,Cao:2014efa,Cahill-Rowley:2014ora}. 
Those analyses found that two regions at a DM mass around $\approx85$~GeV and $\approx 180$~GeV are compatible with the excess. In this case, DM is the lightest neutralino\footnote{Neutralinos are fermionic partner particles of the Standard Model bosonic fields (i.e. the B, W and Higgs fields). Corresponding to their composition the neutralinos can be binos (partners of B-field),
winos (partners of W fields), higgsinos (Higgs partners) or a combination of these quantum states.} and is a mixture which is dominantly bino and has subdominant wino and/or higgsino components. 

The pMSSM parameter regions consistent with the
GCE lead to novel DM  signals at the Large Hadron Collider (LHC), not covered by searches for simplified models or other SUSY scenarios~\cite{vanBeekveld:2016hbo}. Interestingly one of the higgsino-bino regions is consistent with one of the lowest fine-tuning values of the electroweak sectors found in the pMSSM\cite{vanBeekveld:2016hug}.

In this work we revisit the GCE interpretation in terms of the pMSSM through a fit to the new {\it Fermi}-LAT data, accounting for the most up-to-date phenomenological constraints. In particular we consider the constraints from direct DM searches in experiments like LUX \cite{Akerib:2016vxi}, PANDAS \cite{Tan:2016zwf}, XENON1T \cite{Aprile:2017iyp} and PICO \cite{Amole:2017dex}, as well as from SUSY searches at the LHC. The solutions that we find explain the GCE in the photon energy range from 1 to 10 GeV. To account for the high-energy tail of the GCE, we assume an universal power law as different phenomena have been proposed to explain it based on its tentative X-shaped morphology~\cite{Porter:2017vaa,Horiuchi:2016zwu,Linden:2016rcf}. We will give no further astrophysical interpretation of the high energy tail. 
 
The paper is organized as follows: in section \ref{sec:gce} we describe the status of the GCE as revealed with Pass 8 data, while in Section \ref{sec:analysis_setup} we describe the details of the present analysis. In  section \ref{sec:results} we show the results and discuss the prospects of the pMSSM scenario in terms of LHC run II and direct dark matter searches. Finally, in Section \ref{sec:conclusion} we discuss the implication of the results and present our conclusions.
\newpage
\section{The Galactic Center Excess}
\label{sec:gce}

\begin{figure}[t!]
	\centering
		\includegraphics[width=\textwidth]{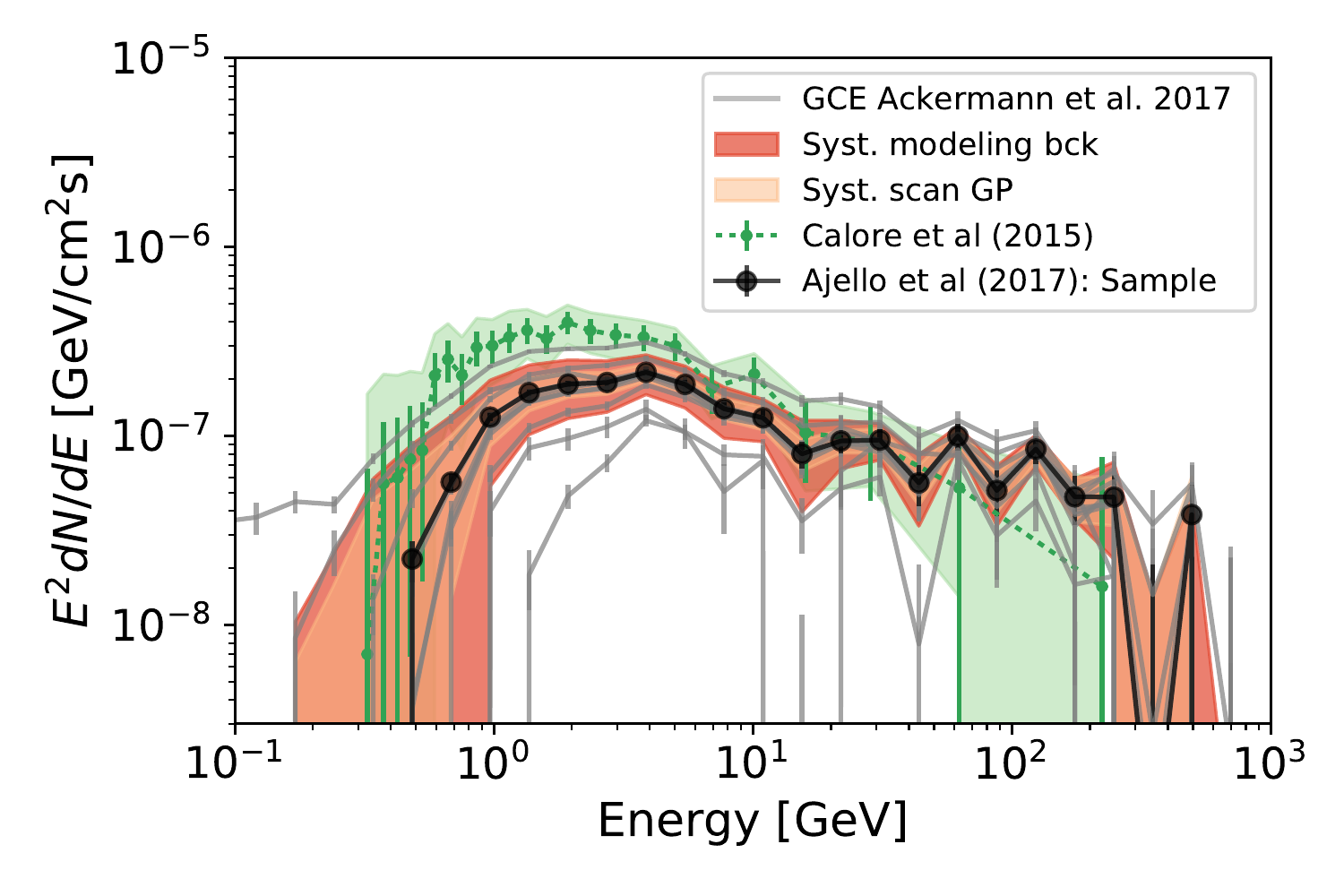}
		\caption{GCE spectra from ref. \cite{1704.03910}. Following ref. \cite{1704.03910} we select the GCE derived using the Sample Model (see section 2.2 of ref. \cite{1704.03910} for description of the model) to perform the fits (black points). The different spectra resulting from using different assumptions for CR production, propagation in the Galaxy and allowing for more freedom in the IC emission fit are shown in gray. The diagonal of two covariance matrices is plotted, one due to excesses found along the Galactic plane in ref. \cite{1704.03910} (light orange band), and one due to the set of all GCE spectra plotted in gray (dark orange band). For comparison the GCE spectrum as found in ref. \cite{Calore:2014xka} (green points) is plotted together with the diagonal of the covariance matrix derived there (light green band).}
		\label{fig:GCE_paper_fermi}
\end{figure}

\begin{figure}[t!]
	\centering
	\begin{subfigure}{.45\textwidth}
		\centering
		\includegraphics[width=\textwidth]{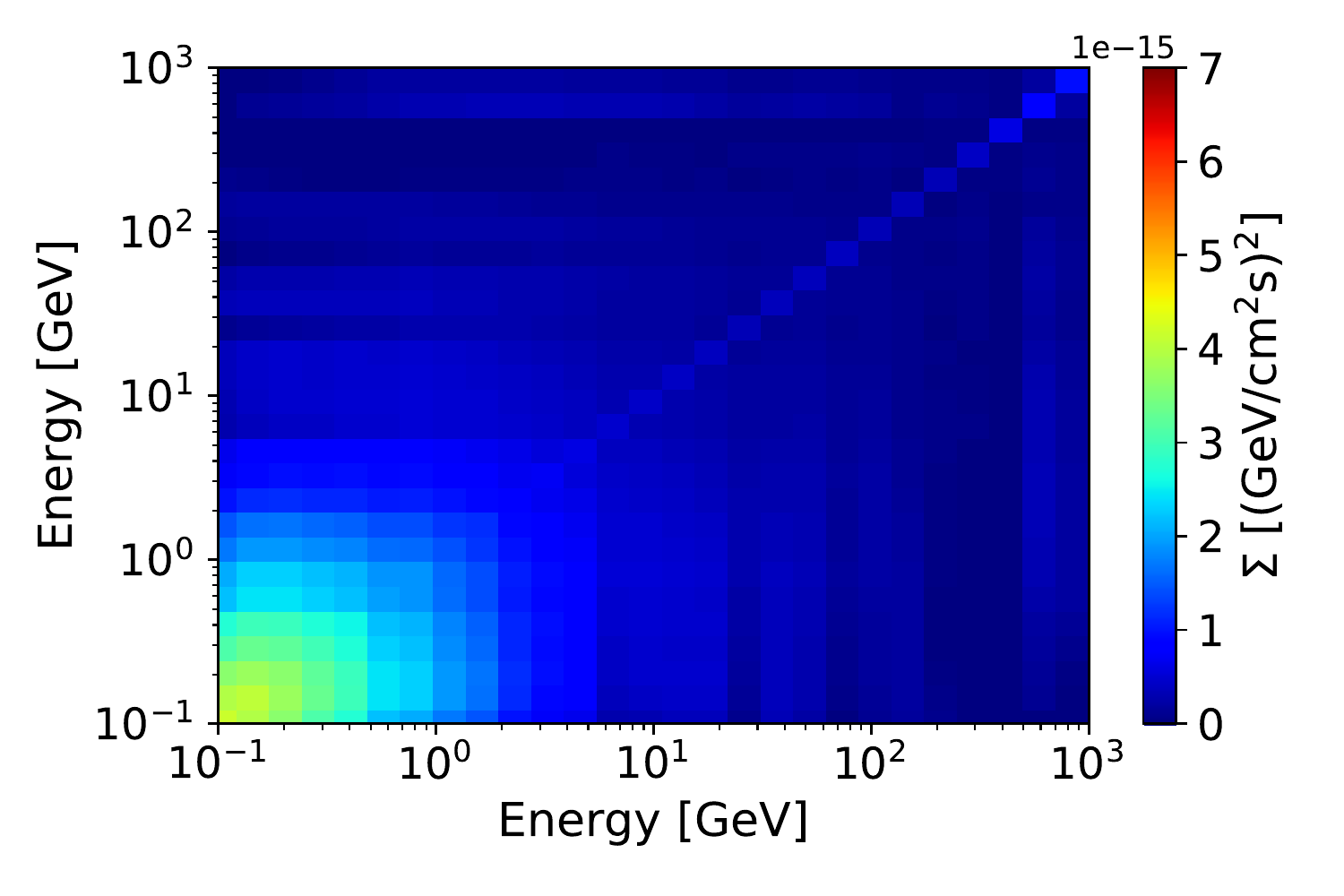}
		\caption{}
		\label{fig:cov_mat_GP}
	\end{subfigure}
	\begin{subfigure}{.45\textwidth}
		\centering
		\includegraphics[width=\textwidth]{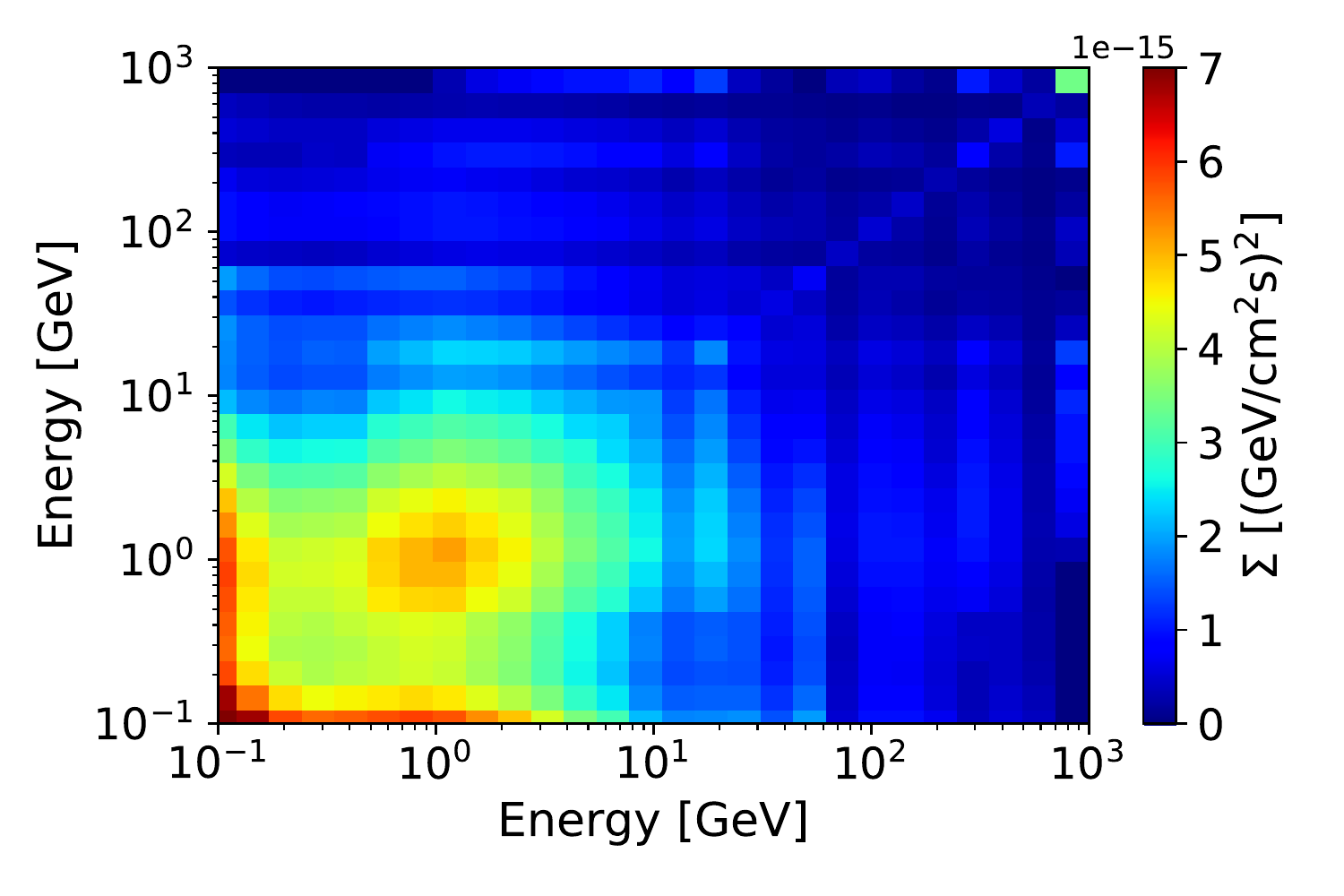}
		\caption{}
		\label{fig:cov_mat_some}
	\end{subfigure}
	\begin{subfigure}{.45\textwidth}
		\centering
		\includegraphics[width=\textwidth]{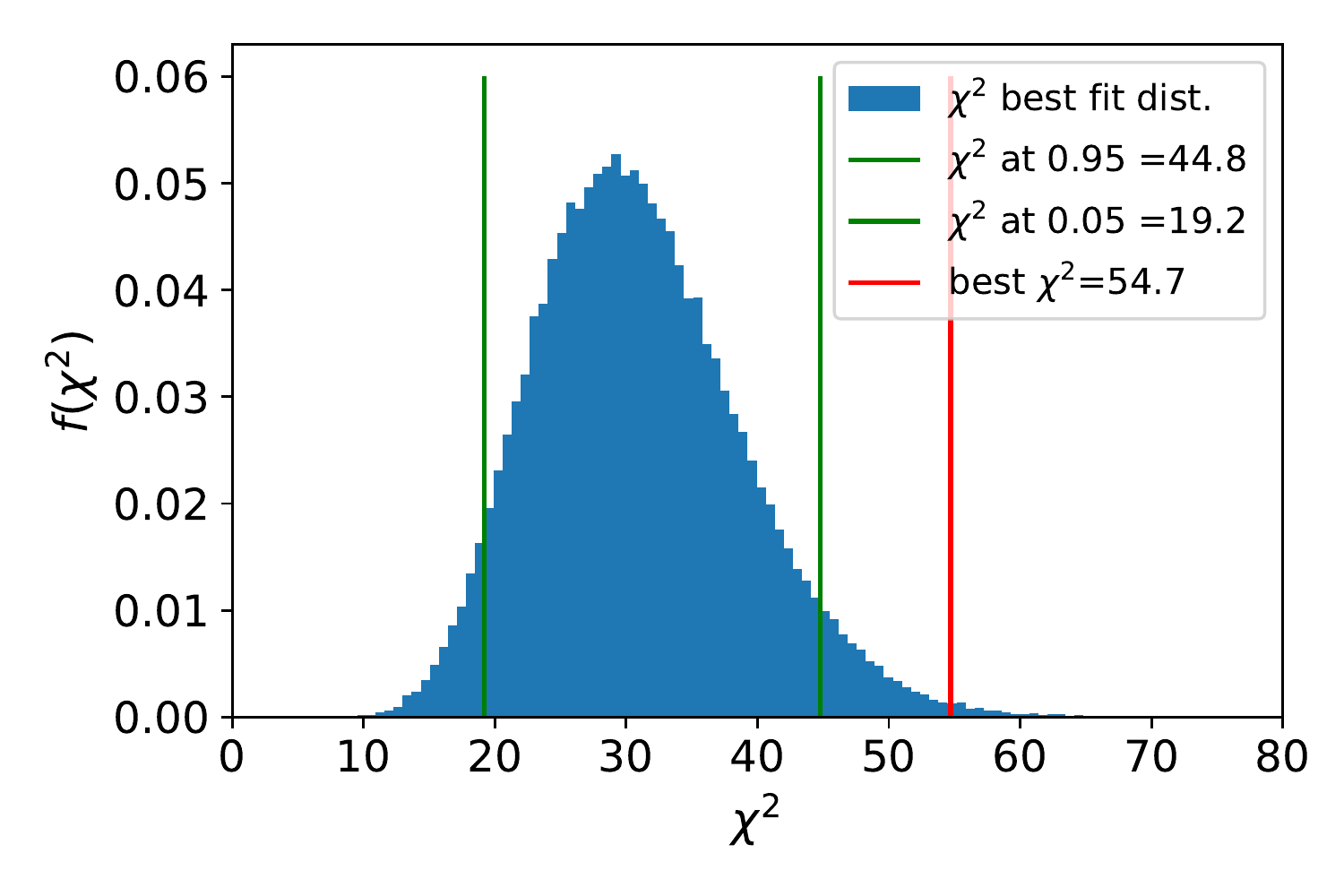}
		\caption{}
		\label{fig:fit_cov_mat_GP}
	\end{subfigure}
	\begin{subfigure}{.45\textwidth}
		\centering
		\includegraphics[width=\textwidth]{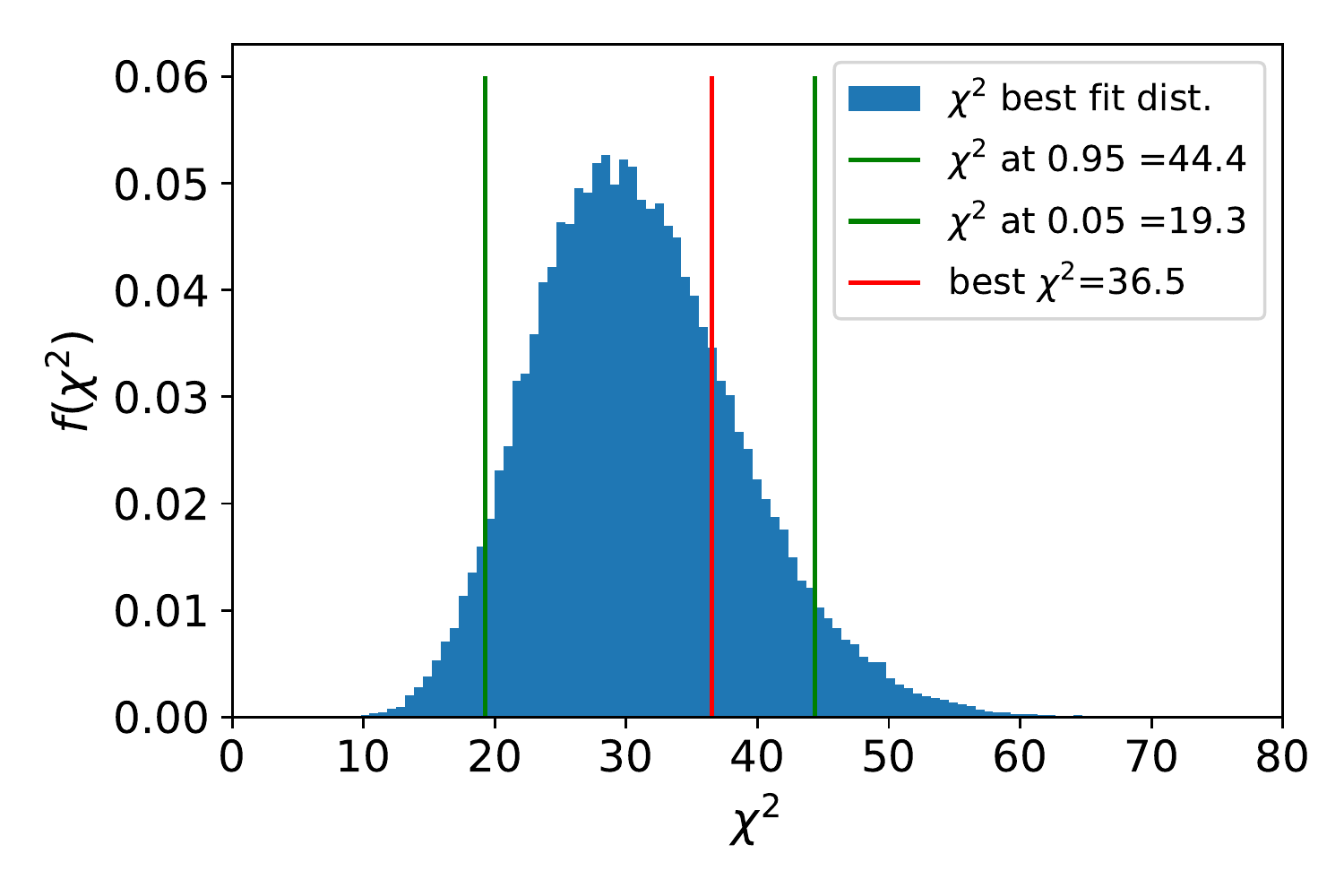}
		\caption{}
		\label{fig:fit_cov_mat_some}
	\end{subfigure}
	\label{fig:chi2dist}
	\caption{(a) Covariance matrix using statistical uncertainties and the spectra from the analysis along the Galactic plane in ref. \cite{1704.03910}. (b) Covariance matrix using the GCE spectra shown by the gray lines in figure \ref{fig:GCE_paper_fermi} plus the statistical uncertainties. (c) Results of scenario 1: the $\chi^2$ distribution of the best fit using the covariance matrix in (a) to fit the toy model of equation \ref{eq:toy}. (d) Results of scenario 2: the $\chi^2$ distribution of the best fit using the composed covariance matrix of (a) and (b) to fit the toy model of equation \ref{eq:toy}.}
\end{figure}

Our analysis uses the GCE measurements presented in ref. \cite{1704.03910}, whose analysis significantly benefits from Pass 8 event-level data due to its improvement in acceptance, the reconstruction of arrival direction and energy, and the sub-selection of events based on the quality of the direction reconstruction. The GCE spectrum is shown in figure \ref{fig:GCE_paper_fermi}. In ref. \cite{1704.03910} the following effects of the systematic uncertainties affecting the properties of the GCE were investigated:

\begin{enumerate}\label{syst}
\item  Changes in the {\it Fermi}-LAT event selection and analysis region.
\item  Using different assumptions for cosmic ray (CR) production and propagation in the Galaxy (GALPROP parameters \cite{Moskalenko:1997gh,Strong:1998fr}), and allowing for more freedom in the fit to inverse Compton (IC) emission \cite{2000ApJ...528..357M} (both indicated by the gray lines in figure \ref{fig:GCE_paper_fermi}).
\item The inclusion of CR sources in the GC region that can induce gamma-ray emission 
\item  The use of alternative distributions of interstellar gas in the GC region.
\item The extension of the {\it Fermi} bubbles to the GC with a data driven method designed to quantify its role in the GCE.
\item As the point sources near the GC region strongly depend on the IEM used to describe the diffuse emission, different point source catalogs based on different IEMs and analyses are tested.
\end{enumerate}
In all the scenarios listed above the GCE is found to be statistically significant. However, the GCE spectral shape varies in the scenarios tested in the following way: at photon energies between 1 and 3 GeV, the flux changes by a factor of $\sim 3$. Above 10 GeV the change is significantly larger, even compatible with zero flux for some background models (gray lines in figure \ref{fig:GCE_paper_fermi}) \cite{1704.03910}. In order to consider the systematic uncertainties explored in ref. \cite{1704.03910} we create {\it two covariance matrices} to capture the effects on the GCE due to:
\begin{itemize}
\item The excesses found along the Galactic plane, see figure \ref{fig:cov_mat_GP}.
\item The variations in the GCE due to item 2 listed above\footnote{We restrict ourselves to using these GCE spectra as the other possibilities explored in ref. \cite{1704.03910} include astrophysical sources that absorb part of the GCE emission. In this work we model the GCE with two different components to account for these effects.}, see figure \ref{fig:cov_mat_some}.
\end{itemize}
To compute the covariance matrices we use (section 4.2.2 of ref. \cite{Calore:2014xka}):

\begin{equation}\label{eq:cov}
\Sigma_{i,j} = \left\langle \Phi_i\Phi_j \right\rangle -\left\langle \Phi_i \right\rangle \left\langle \Phi_j \right\rangle.
\end{equation}
Here, $\Phi_i$ represents the measured gamma-ray flux at the $i-$th energy bin (the black points in figure \ref{fig:GCE_paper_fermi}). The average is over the different spectra under consideration to build the covariance matrix. To fit a particular model $\Phi^m(\vec{\theta})$ we find the set of $\vec{\theta}$ parameters that minimizes the $\chi^2$ function:
\begin{equation}
\chi^2 = \sum_{i,j}{(\Phi^d_i - \Phi^m_i)\hat{\Sigma}_{ij}^{-1}(\Phi^d_j - \Phi^m_j)},
\end{equation}
where $\Phi^d_i$ represents the GCE flux in energy bin $i$. The inverse of the covariance matrix is denoted by $\hat{\Sigma}^{-1}$. To test if a model $\Phi^m(\vec{\theta})$ is rejected as an explanation of the data $ \{\Phi^d, \hat{\Sigma}\} $ we cannot use a reduced $\chi^2$ to compute p-values as the models normally used to fit the GCE are not linear and the data we aim to fit is already obtained from a fit\footnote{See ref. \cite{2010arXiv1012.3754A} for more information.}. Therefore we use the following method:
\begin{enumerate}
	\item Find the set of $\vec{\theta}_{best}$ that minimize $\chi^2$ for a particular $\{\Phi^d, \hat{\Sigma}\} $ and save $\chi^2_{best}$.
	\item Create a set of 100.000 pseudo-random data normal distributed with mean at $\Phi^m(\vec{\theta}_{best})$ according to $\hat{\Sigma}$.
	\item Compute $\chi^2$ between $\Phi^m(\vec{\theta}_{best})$ and each one of the 100.000 pseudo-random data created in 2.
	\item Create a $\chi^2$ distribution using the values from item 3 and find the values $\chi^2_{5\%}$ and $\chi^2_{95\%}$ at which the integrated distribution covers 5\% and 95\% of the total $\chi^2$ distribution, respectively.
	\item If the $\chi^2_{best}$ found in 1 is lower than $\chi^2_{5\%}$ or greater than $\chi^2_{95\%}$, the $\Phi^m(\vec{\theta}_{best})$ is rejected as an explanation of the data  $\{\Phi^d, \hat{\Sigma}\} $ at 95\% CL.
	\item If the $\chi^2_{best}$ found in 1 is between  $\chi^2_{5\%}$ and $\chi^2_{95\%}$, the $\Phi^m(\vec{\theta}_{best})$ {\it cannot} be rejected as an explanation of the data  $\{\Phi^d, \hat{\Sigma}\} $.
\end{enumerate}

In the following we present two examples where we test a toy model to explain the GCE using this method. Our toy model is composed of two power laws with exponential cut-off \footnote{A power law with an exponential cuf-off is a typical form for the flux originating from astrophysical sources.}:

\begin{equation}\label{eq:toy}
\Phi^m_{toy} = \sum_{a=1,2} {N_a} \left(\frac{E}{E_{0,a}}\right)^{-(\alpha_a - \beta_a\log{E/E_{0,a}})}  e^{\left(\frac{E-E_{0,a}}{E_{cut,a}}\right)},
\end{equation} 

with $N_a$ two normalizations at reference energies $E_{0,a}$ and with $E_{cut,a}$ the energies of the cut offs. 

\textbf{Scenario 1. Covariance matrix from GP spectra.}\\
First we apply the method using the covariance matrix visualized by figure \ref{fig:cov_mat_GP}, where only the excesses along the Galactic plane plus the statistical uncertainties are considered. The results are presented in figure \ref{fig:fit_cov_mat_GP}, where the $\chi^2_{best}$ found tells us that the toy model is rejected as an explanation of the GCE. We expected this result as the toy model looks similar to the DM emission plus {\it Fermi}-bubbles component tested in ref. \cite{1704.03910} and, as already mentioned, that model is found to provide a poor fit to the GCE when considering the excesses along the Galactic plane. 

\textbf{Scenario 2. Covariance matrix considering the GCE spectral variations\footnote{This is the covariance matrix used in this work.}.}\\
For the second example we apply the method to the same toy model, but instead use a covariance matrix composed of the statistical uncertainties, the systematics associated with the excesses in the Galactic plane (figure \ref{fig:cov_mat_GP}) and the variations in the GCE from modeling the Galactic interstellar emission (figure \ref{fig:cov_mat_some}). Results are presented in figure \ref{fig:fit_cov_mat_some}. In this case we find that the toy model {can not} be rejected as an explanation of the GCE.\\
Note that in both scenarios, the GCE is significant. \\

Regarding the DM interpretation of the GCE, the authors of ref. \cite{1704.03910} conclude that the spectrum and morphology of the GCE are not evidently consistent with expectations from DM annihilation as DM-like signals observed in other regions of the Galactic plane, where such signals are not expected, give a handle on the magnitude of systematic uncertainties due to diffuse emission modeling in the Galactic Center. However, as discussed in ref. \cite{1704.03910},  one can not exclude the possibility that the GCE is the result of other gamma-ray sources: for instance a known astrophysical emitter (e.g. Fermi bubbles, millisecond pulsars) on top of a DM-induced component. To explore this possibility we model the GCE spectrum with a component from WIMP annihilation together with a generic astrophysical component. The astrophysical component is modeled as a power law in photon energy with an exponential cut off, like the one shown in equation \ref{eq:toy}. The other power law in the toy model is replaced by SUSY WIMP predictions. We will follow the same procedure to model the uncertainties as in scenario 2. In addition to the uncertainties of the {\it Fermi}-LAT, we add 10\% uncorrelated uncertainty to account for uncertainties in the Monte Carlo event generators that stem from parton showering models and its model parameters \cite{Agrawal:2014oha}. This is achieved by substituting $\hat{\Sigma}^{-1}_{ij} \rightarrow \hat{\Sigma}^{-1}_{ij}+\delta_{ij}\left(\Phi^m_i\right)^2\sigma^2$, where $\sigma=10\%$ \cite{Caron:2015wda}. This will be referred to as the $100\%$ flux scenario. \\
As was pointed out in section \ref{sec:introduction}, point sources are likely to account for a significant fraction of the GCE. However, sources too dim to be detected individually  collectively produce diffuse emission. To take this effect into account, we also explore a hybrid model. Here we assume that gamma-ray emitters of astrophysical origin reduce the GCE flux to 40\% of the total flux while keeping the shape intact. The remaining 40\% of the total flux is fitted by dark matter annihilation and the one free power law as in the $100\%$ flux scenario. In this hybrid model the uncertainties are downscaled to 40\% as well. This will be referred to as the $40\%$ flux scenario. \\
We have chosen the value of $40\%$, since the no-DM-hypothesis in this case gives a p-value of $0.056$\footnote{The no-DM-hypothesis for the 100$\%$ flux scenario amounts to a p-value of $10^{-16}$.}. If we reduce the flux even more, the remaining GCE is not significant anymore. The reduced flux shows the effect of the GCE explained by a reduced MSSM DM component on the pMSSM parameter space.


\section{Models and Constraints}\label{sec:analysis_setup}
The phenomenological MSSM (pMSSM) \cite{Djouadi:1998di} is defined by imposing the following constraints on the MSSM:
\begin{itemize}
\item One assumes degenerate first and second generation squark and slepton masses. 
\item All trilinear couplings of the first and second generation sfermions are set to zero.
\item There are {\em no} new sources of CP violation.
\item One demands minimal flavor violation, so all sfermion mass matrices are assumed to be diagonal.
\end{itemize}
Applying these conditions one ends up with a 19-dimensional model that can be  parametrized  as follows: the sfermion soft-masses are described by the first and second generation squark masses $m_{\tilde Q}$, $m_{\tilde{U}_1}$ and $m_{\tilde{D}_1}$, the third generation squark masses $m_{\tilde{Q}_3}$, $m_{\tilde{U}_3}$ and $m_{\tilde{D}_3}$, the first and second generation of slepton masses $m_{\tilde L}$, $m_{\tilde{E}}$ and the third generation of slepton masses $m_{\tilde{L}_3}$, $m_{\tilde{E}_3}$. The trilinear couplings of the
third generation of sfermions $A_{\tilde t}$, $A_{\tilde b}$ and $A_{\tilde \tau}$ are assumed to be non-zero. The Higgs sector is  described by the ratio of the Higgs vacuum expectation values tan $\beta$ and the  soft Higgs masses $m^2_{H_{u,d}}$. Instead of these Higgs masses, we choose to use the higgsino mass parameter $\mu$ and the mass of the pseudoscalar Higgs $m_A$ as input parameters. Finally one chooses the gaugino masses $M_1$, $M_2$ and $M_3$. 

It is nontrivial to scan a large parameter space and apply all constraints stemming from various astrophysical and particle physics experiments \cite{Berger:2008cq}. In this paper we use the following strategy. We use the fit points of ref. \cite{Caron:2015wda} and ref. \cite{vanBeekveld:2016hug} as seeds to calculate around 4 million new
parameter set evaluations (points).  In an iterative procedure the best-fit points of the foregoing iteration are used as additional seeds to sample new model points, where a truncated multi-dimensional Gaussian distribution is used around each parameter of the seed to sample new points \cite{1232326}. The width of the Gaussian distribution is chosen to be 0.5 times the value of the seed point in each dimension. 

SUSPECT \cite{Djouadi:2002ze} is used as spectrum generator, while the Higgs mass is calculated using FeynHiggs \cite{Bahl:2016brp, Hahn:2013ria,Frank:2006yh,Degrassi:2002fi, Heinemeyer:1998yj}. MicrOMEGAs 4.3.2 \cite{Barducci:2016pcb} is used to compute flavor variables, $g-2$, $\Omega_{\rm DM} h^2$, the velocity weighted annihilation cross section and the spin-dependent and spin-independent WIMP-nucleon scattering cross sections ($\sigma_{\rm SD}$ and $\sigma_{\rm SI}$). The gamma ray spectrum is computed using DarkSUSY 5.1.3 \cite{DarkSUSY}.

We require the value of observables,  as calculated for the model parameters, to lie within the $2\sigma$ interval around the experimentally obtained value, unless indicated otherwise. The following limits are applied to the model points:
\begin{itemize}
\item LEP limits on the masses of the lightest chargino ($m_{\tilde{\chi}^{\pm}_1}~>~103.5$~GeV) and sleptons ($m_{\tilde{l}}>90$~GeV)~\cite{LEP:working}.
\item Constraints on the invisible and total width of the $Z$-boson,  $\Gamma_{Z, {\rm inv}}~=~499.0~\pm~1.5$~MeV and $\Gamma_{Z}~=~2.4952~\pm~0.0023$~GeV respectively, obtained from $Z$-pole measurements at LEP \cite{Carena:2003aj}.
\item The LHC measurements of the Higgs boson mass \cite{Aad:2014aba, Cao:2014efa}. On top of this we account for a theoretical SUSY uncertainty of 3 GeV, selecting models with a Higgs boson within the mass range of $122$ GeV $\leq m_{h_0} \leq 128$ GeV.
\item An upper bound of the muon anomalous magnetic dipole moment $\Delta(g-2)_{\mu} < 40\times 10^{-10}$, taking into account the fact that the SM prediction lies well outside the experimentally obtained value: $(24.9\pm 6.3)\times 10^{-10}$ \cite{Roberts:2010cj}.
\item Measurements of the $B/D$-meson branching fractions: Br$(B_{(s)}^0~\rightarrow~\mu^+\mu^-)$~\cite{Aaij:2013aka}, Br$(\bar{B}~\rightarrow~X_s~\gamma)$~\cite{Misiak:2015xwa, Czakon:2015exa}, Br($B^+~\rightarrow~\tau^+~\nu_{\tau}$)~\cite{Kronenbitter:2015kls}, Br$(D_s^+~\rightarrow~\mu^+~\nu_{\mu})$~\cite{Widhalm:2007ws} and Br($D^+_s~\rightarrow~\tau^+~\nu_{\tau})$~\cite{Onyisi:2009th}. 
\item  Results of (heavy) Higgs searches at LEP, the Tevatron and the LHC as implemented in HiggsBounds 4.3.1 \cite{Bechtle:2015pma}. 
\item A determination of the exclusion of a model point using SUSY-AI (with the 13 TeV center-of-mass option). SUSY-AI is a machine learning tool, trained with ATLAS data, which is able to exclude model points in the pMSSM parameter space based on ATLAS results \cite{Caron:2016hib, Barr:2016sho}.  
\item Limits set by ATLAS on stop production in simplified MSSM scenarios using 2016 ATLAS data, which are not yet included into SUSY-AI. Depending on the stop decay, models are excluded if they fall in the excluded region of the neutralino-stop plane for $\tilde{t} \rightarrow W b \tilde{\chi}^0_1$ \cite{ATLAS:2016xcm}, $\tilde{t} \rightarrow c \tilde{\chi}^0_1$ \cite{Aaboud:2016tnv,Aad:2014nra} and $\tilde{t} \rightarrow \tilde{\chi}^{+}_1 b$ \cite{Sirunyan:2017wif}.
\item Constraints on the WIMP-nucleus scattering cross section from various detector experiments, using DDCalc \cite{ddcalc:2017lvb} with the 2016 results from LUX \cite{Akerib:2016vxi}, the 2017 limits from PICO \cite{Amole:2017dex}, the 2016 limits from PandaX\cite{Tan:2016zwf} and the recent XENON1T limits \cite{Aprile:2017iyp}. As in ref. \cite{Aad:2015baa}, we reject models that are excluded by LUX, PICO, PandaX or XENON1T with more than $3\sigma$ to account for the form factor and astrophysical uncertainties. In the regions found in this analysis the limits set by PICO are stronger than the limits set by IceCube \cite{Aartsen:2016zhm}. Therefore the limits of IceCube are not used in this analysis. 
\item The uncertainties of the amount of dark matter in the line-of-sight (the $J$-factor) are as in ref. \cite{Calore:2014nla}. 
\item Limits on the velocity weighted annihilation cross section obtained by analyzing dwarf galaxies \cite{Albert:2005kh,Ackermann:2015zua,Ahnen:2016qkx}, if the dominant annihilation channel is mainly $W^+W^-$.
\end{itemize}

\section{Results}
\label{sec:results}
\begin{figure}[t!]
\centering
\begin{subfigure}{.5\textwidth}
  \centering
  \includegraphics[width=\textwidth]{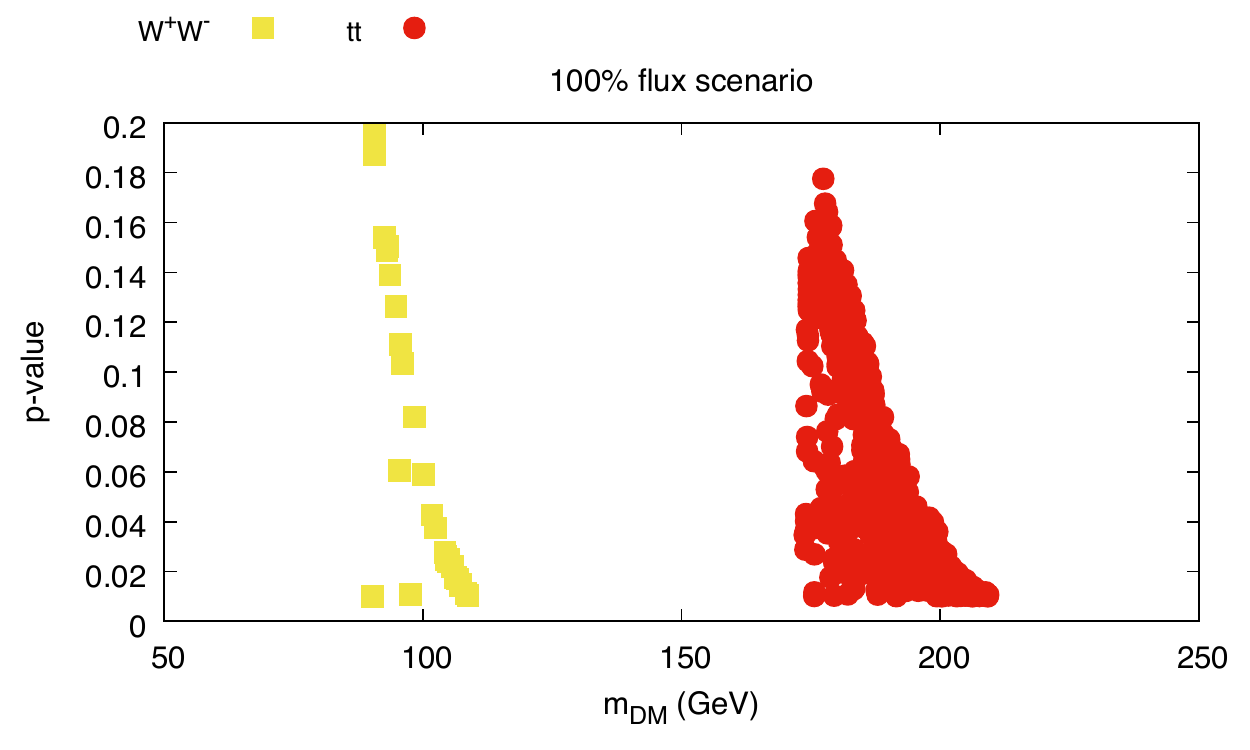}
\end{subfigure}%
\begin{subfigure}{.5\textwidth}
  \centering
  \includegraphics[width=\textwidth]{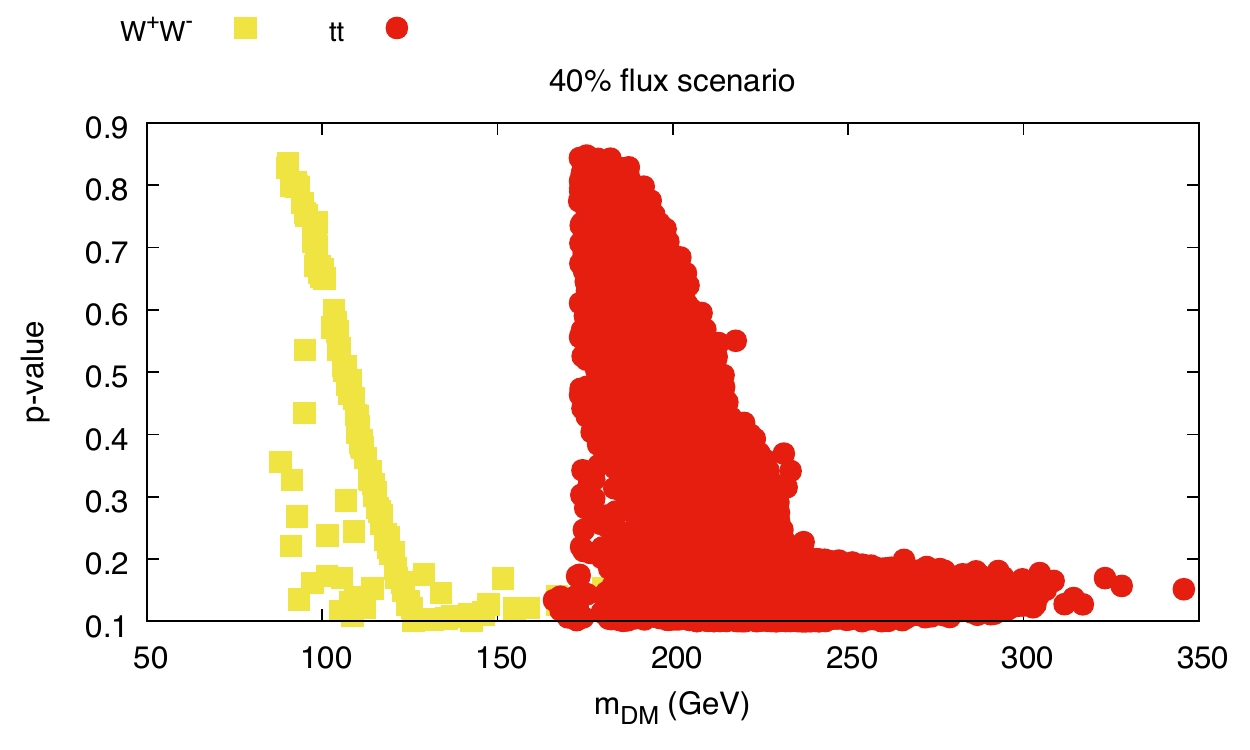}
\end{subfigure}
\caption{P-value as a function of the dark matter mass for the 100$\%$ DM assumption (left figure) and the 40$\%$ DM assumption (right figure).} 
\label{fig:pvalues}
\end{figure}

\begin{figure}[b]
	\centering
	\begin{subfigure}{.5\textwidth}
		\centering
		\includegraphics[width=\textwidth]{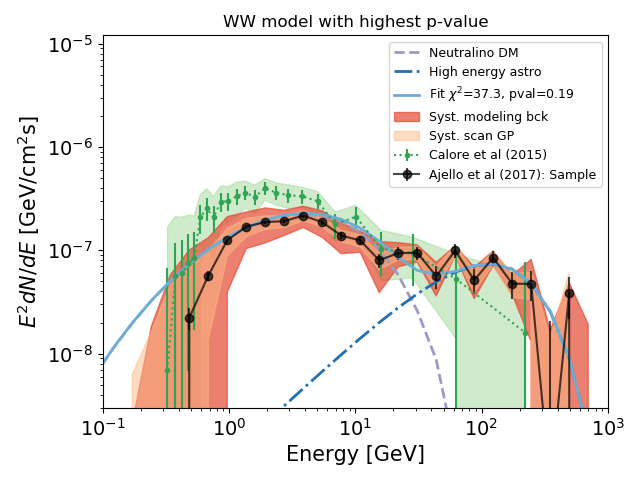}
	\end{subfigure}%
	\begin{subfigure}{.5\textwidth}
		\centering
		\includegraphics[width=\textwidth]{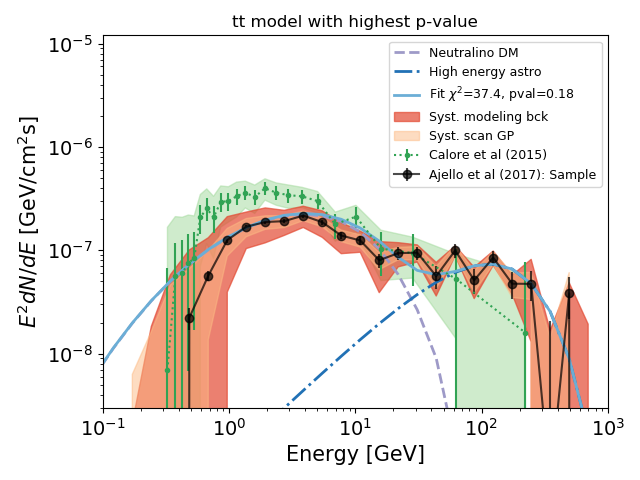}
	\end{subfigure}
    \caption{Spectral energy distribution of the $100\%$ DM models with the highest p-value. The left and right panel correspond to $WW/tt$ type of solutions.}
	\label{fig:photon-spectra}
\end{figure}

Figure \ref{fig:pvalues} shows the p-value of the fit as a function of the DM particle ($\tilde{\chi}^0_1$). Two solutions are visible. For a DM mass between
$80-120$~GeV, DM annihilates predominantly to a pair of W bosons ($W^+W^-$ region).
Here the best p-value is 0.19 for the 100$\%$ flux scenario and 0.83 for the 40$\%$ flux scenario. If the neutralino mass is heavier, the photon spectrum is shifted to higher energies and the fit to the GC excess is worse, as clearly visible in the figure. 
DM with a mass of 175-220 GeV annihilates predominantly to a pair of top quarks ($t\bar{t}$ region). The best p-value for this region is 0.12 for WW the 100$\%$ flux scenario and 0.84 for tt the 40$\%$ flux scenario. In both cases, the free power law accounts for the tail of the excess, starting at photon energies of 10 GeV. We do not give a further interpretation of the tail of the GCE. We find that, after selecting models that have a Higgs boson in the right mass range and that evade the LEP SUSY mass limit, in particular the dark matter direct detection constraints (both spin-dependent and spin-independent) are important. The other limits listed in section \ref{sec:analysis_setup}, including the limits on the velocity-weighted annihilation cross section, turn out to be of lesser importance.  \\

Both dark matter solutions overlap  with the solutions found in ref. \cite{Caron:2015wda} and ref. \cite{Bertone:2015tza}. 
Figure \ref{fig:photon-spectra} shows the spectrum of the GCE including the systematic uncertainties associated with the galactic diffuse emission modeling, as outlined in section \ref{sec:gce}, together with the spectra predicted by the pMSSM for the points giving the best p-value in the two regions identified in our scan. We will now discuss each of the regions separately.

\begin{figure}[t]
\centering
\begin{subfigure}{.5\textwidth}
  \centering
  \includegraphics[width=\textwidth]{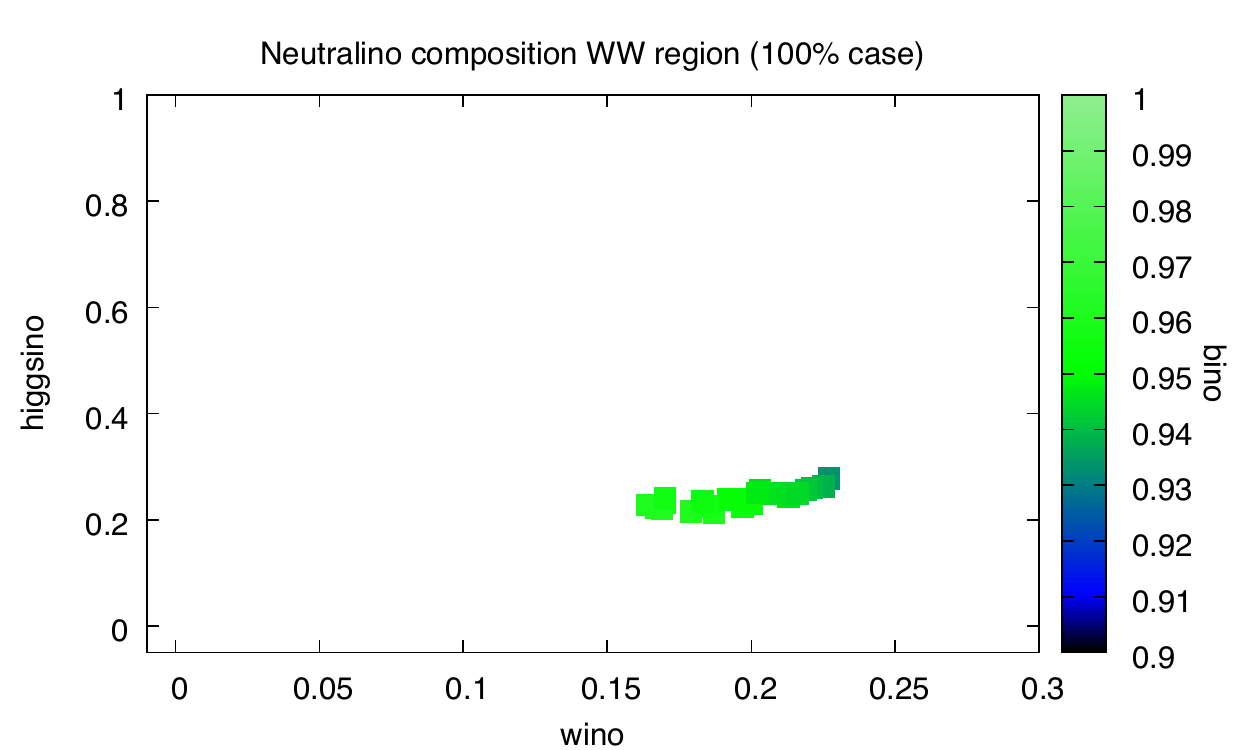}
\end{subfigure}%
\begin{subfigure}{.5\textwidth}
  \centering
  \includegraphics[width=\textwidth]{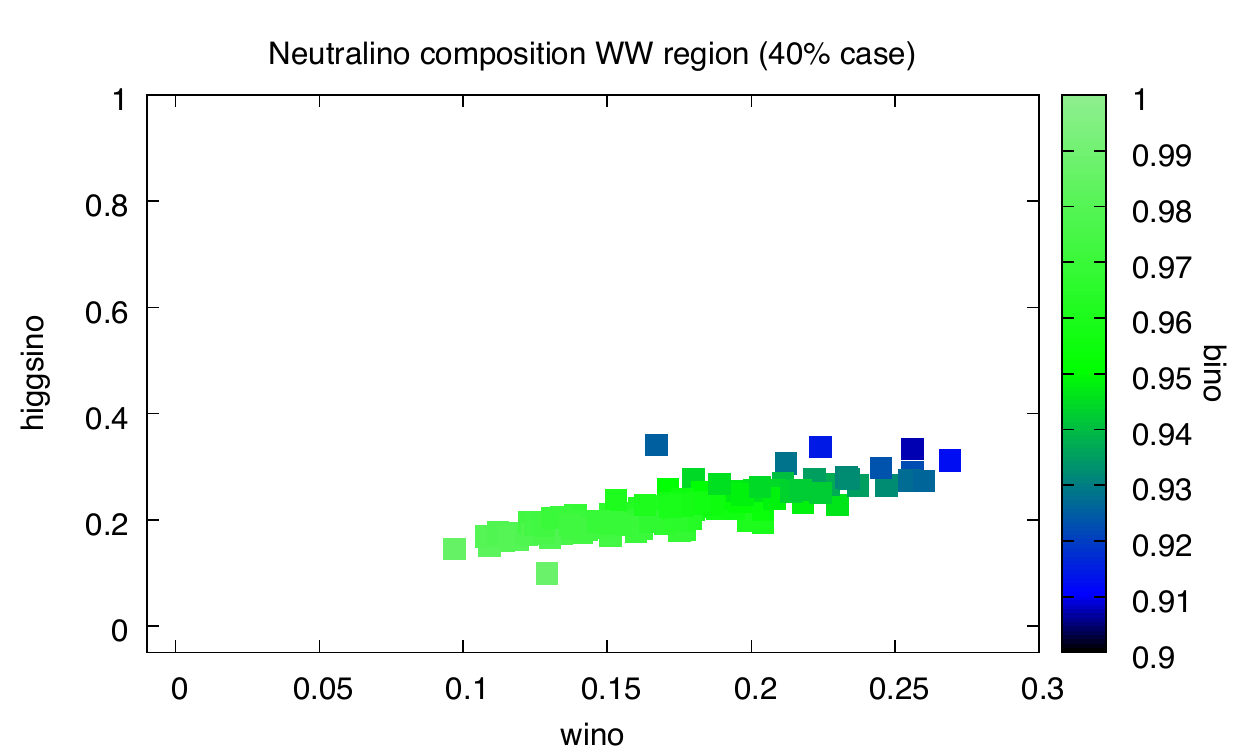}
\end{subfigure}
\begin{subfigure}{.5\textwidth}
  \centering
  \includegraphics[width=\textwidth]{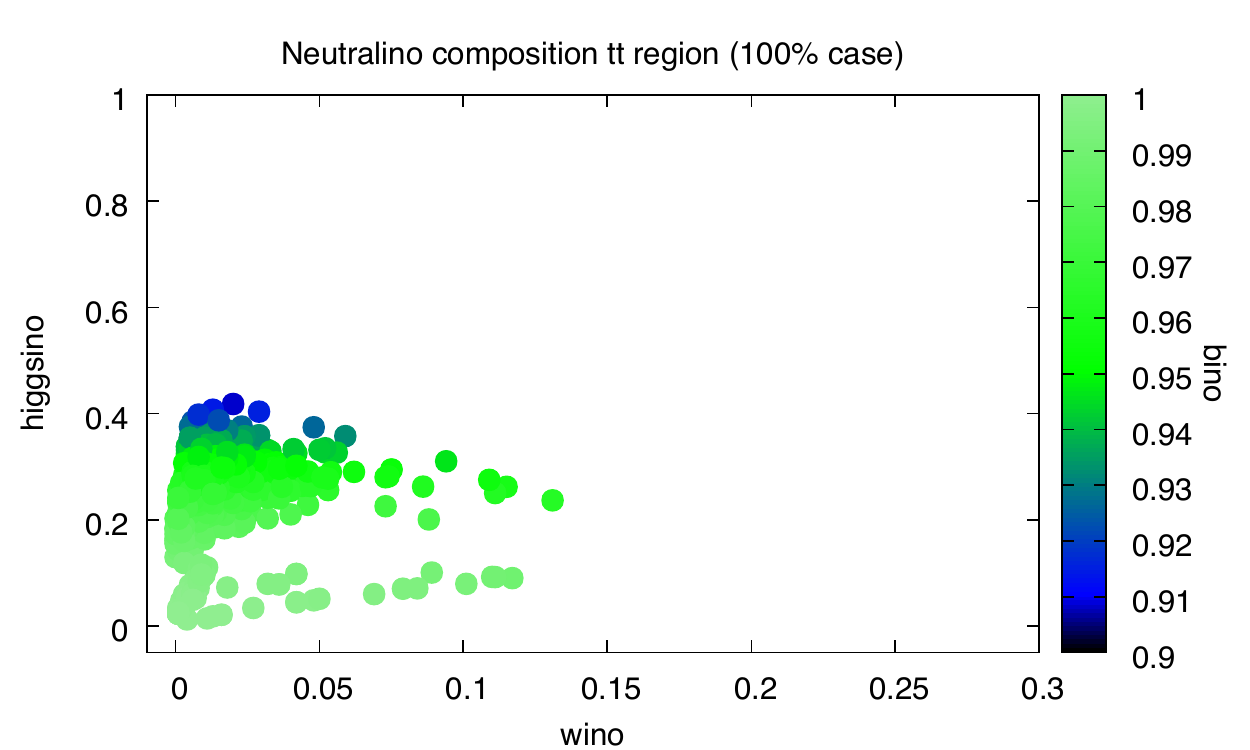}
\end{subfigure}%
\begin{subfigure}{.5\textwidth}
  \centering
  \includegraphics[width=\textwidth]{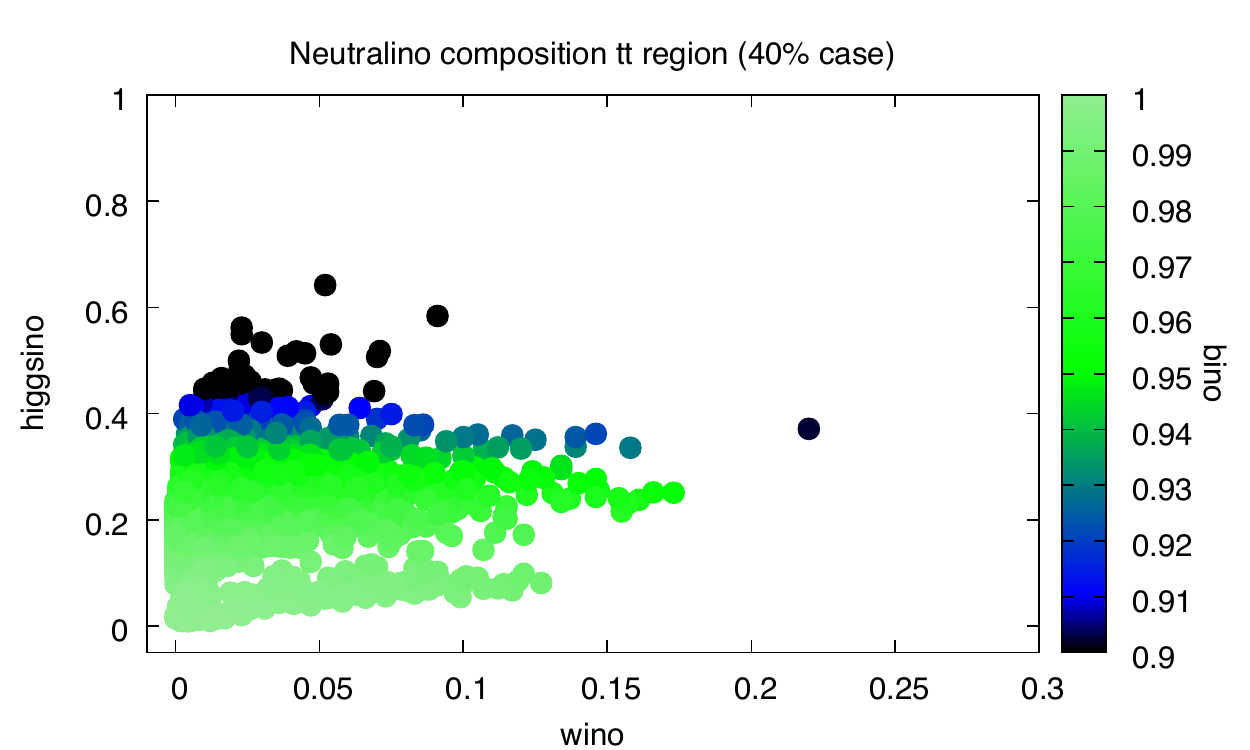}
\end{subfigure}
\caption{Composition of neutralino dark matter where bino $\equiv |N_{11}|$, wino $\equiv |N_{12}|$ and higgsino $\equiv (|N_{13}|^2+|N_{14}|^2)^{1/2}$. Left and right panels represent the 100\% and 40\% flux scenarios respectively, while the top and bottom panels represent the $WW/tt$ solutions.}
\label{fig:wimp_comp}
\end{figure}

\textbf{In the $W^+W^-$ region}, the dominant DM annihilation channel is mediated by a light chargino in the t-channel for present-day annihilation. If the predicted relic density does not make up the measured value of  $\Omega h^2 = 0.118$  \cite{Ade:2013zuv}, the neutralino contributes only a fraction  $\xi = \frac{\Omega h^2_{{\rm model}}}{\Omega h^2_{\rm Planck}}$  to the total Dark Matter.
The photon flux from DM annihilation  then needs to be rescaled by $\xi^2$. The photon flux is proportional to the velocity-weighted present-day annihilation cross-section $\langle \sigma v \rangle$. To allow for a photon flux (and therefore $\xi^2 \langle\sigma v\rangle$)  that is big enough to explain the excess flux, we find that the DM particle must be bino-like with a smaller fraction of either wino, higgsino or both.
Figure \ref{fig:wimp_comp} shows the bino, wino and higgsino composition of the solutions. Larger wino and
higgsino compositions yield a too efficient
early universe annihilation and therefore
rather small values of $\xi$ that is too small to explain the GCE.
A larger higgsino
composition  is also
constrained by  the limits on the spin-dependent DM-nucleus scattering cross section. Adding some wino fraction relaxes these direct detection constraints. 

\begin{figure}[t!]
\centering
\begin{subfigure}{.5\textwidth}
  \centering
  \includegraphics[width=\textwidth]{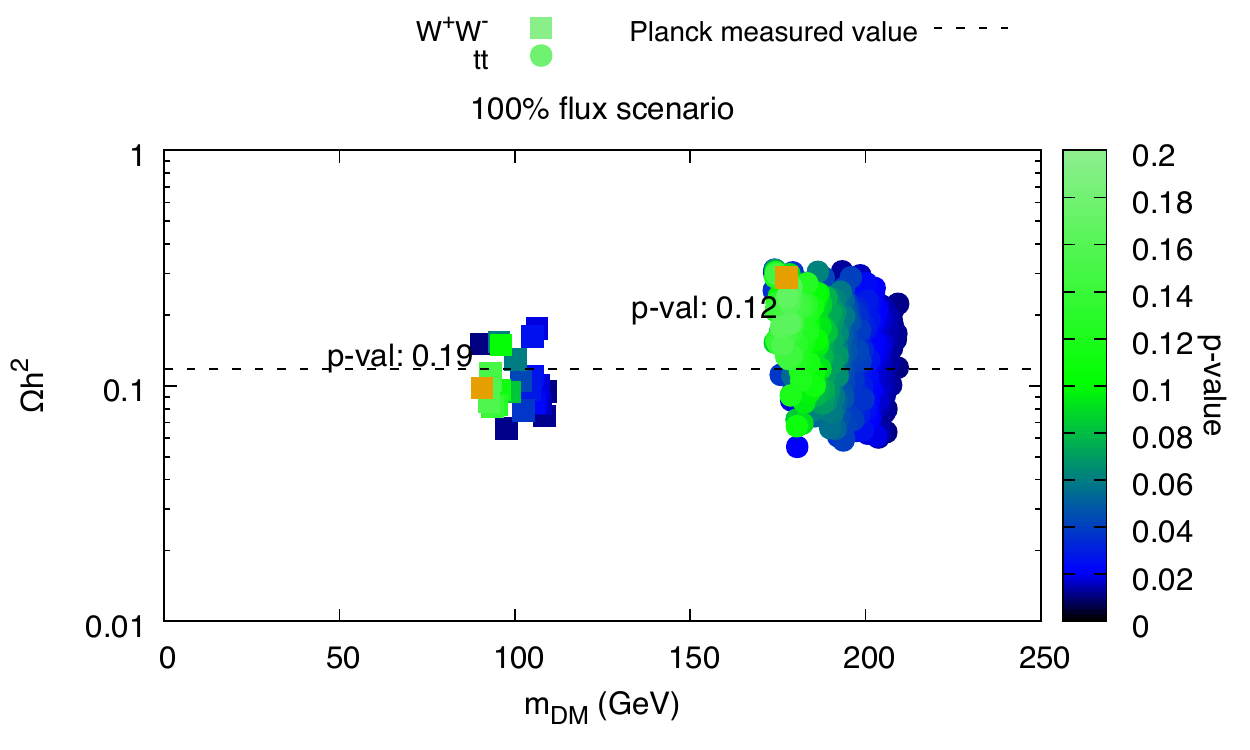}
  \caption{Neutralino mass against $\Omega{\textrm{h}^2}$\\for the 100\% scenario}
  \label{fig:neutralino_oh2}
\end{subfigure}%
\begin{subfigure}{.5\textwidth}
  \centering
  \includegraphics[width=\textwidth]{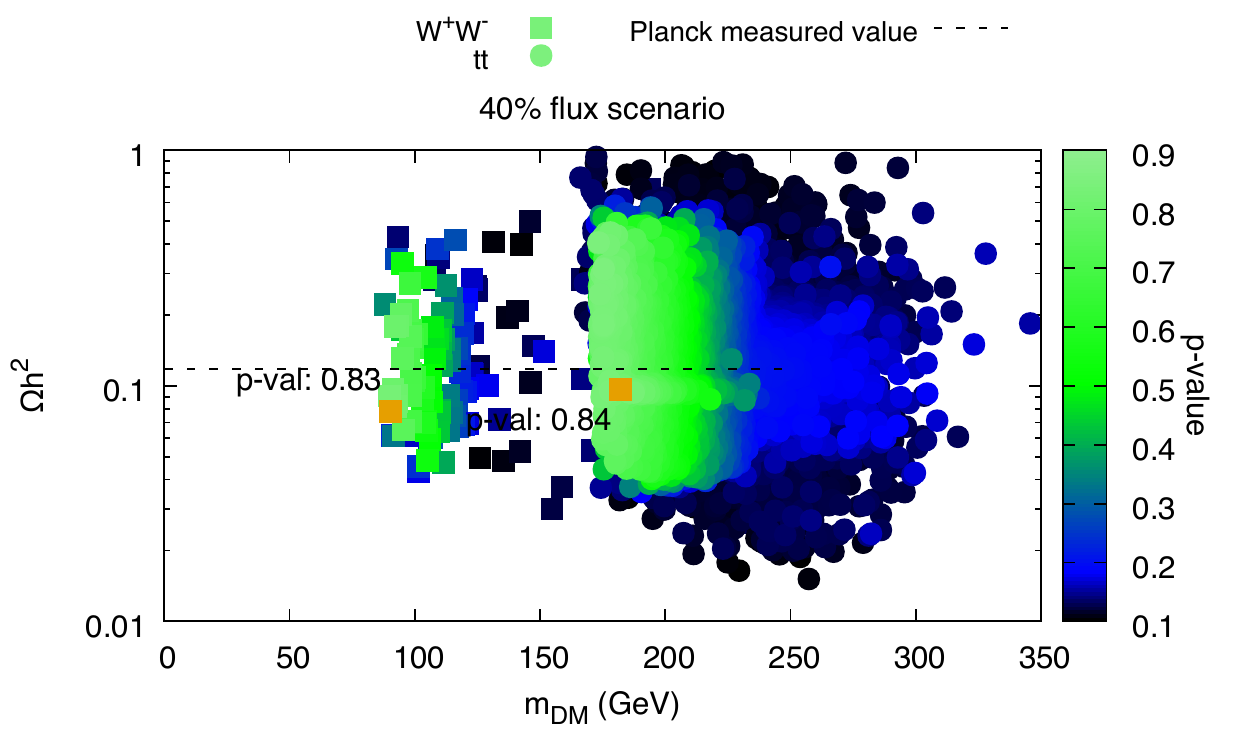}
  \caption{Neutralino mass against $\Omega{\textrm{h}^2}$\\for the 40\% scenario}
  \label{fig:neutralino_oh2_40p}
\end{subfigure}
\begin{subfigure}{.5\textwidth}
  \centering
  \includegraphics[width=\textwidth]{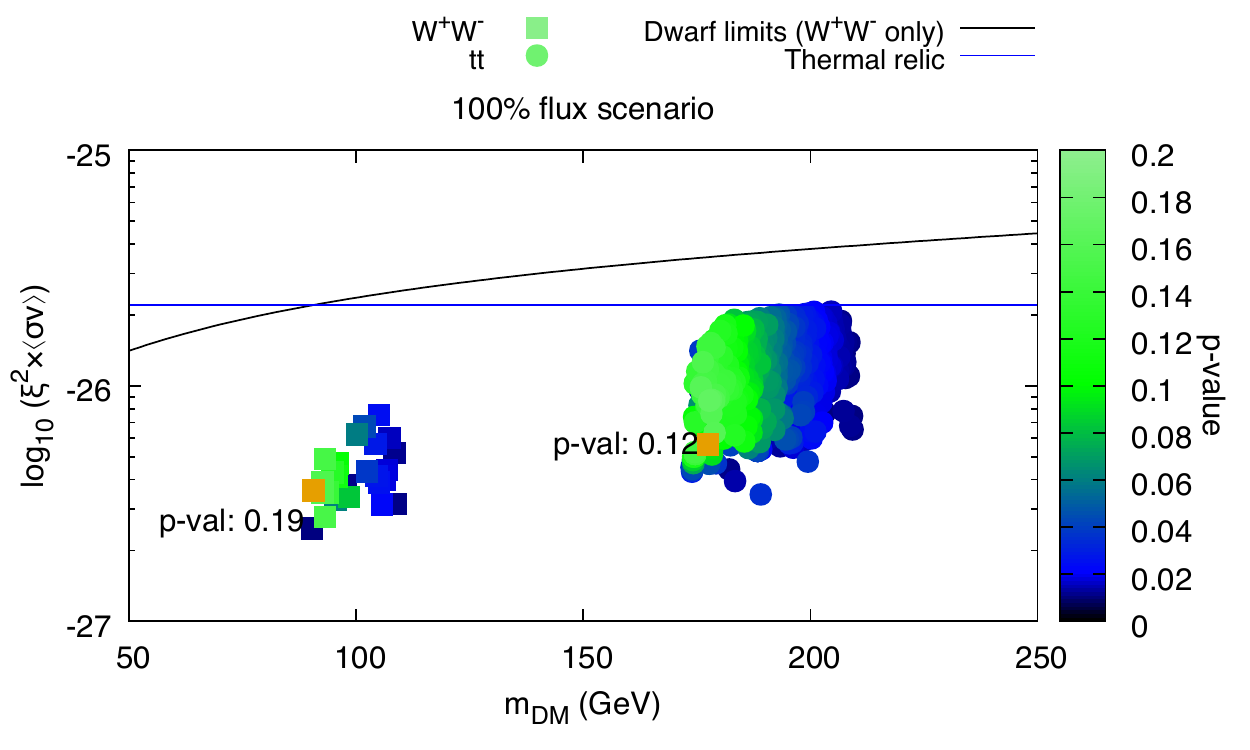}
  \caption{Neutralino mass against $\xi^2\langle\sigma v\rangle$ \\for the 100\% scenario}
  \label{fig:neutralino_sv}
\end{subfigure}%
\begin{subfigure}{.5\textwidth}
  \centering
  \includegraphics[width=\textwidth]{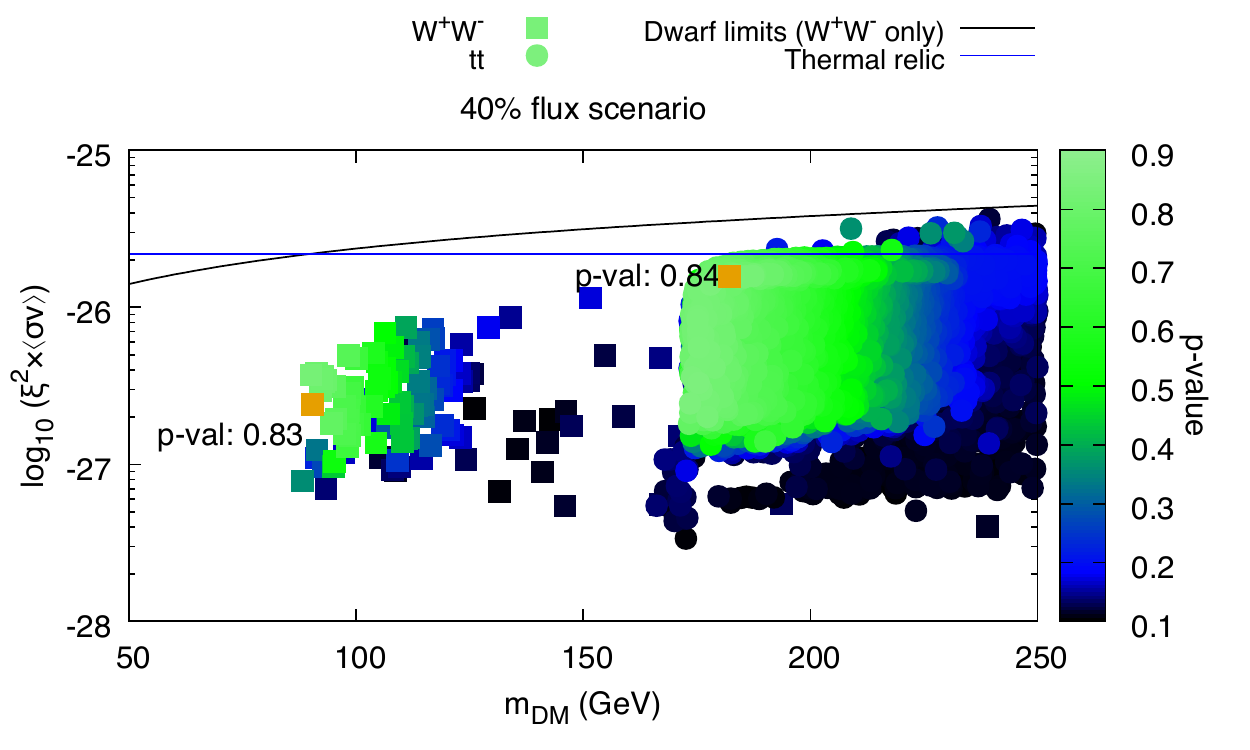}
  \caption{Neutralino mass against $\xi^2\langle\sigma v\rangle$\\for the 40\% scenario}
  \label{fig:neutralino_sv_40p}
\end{subfigure}
\caption{Neutralino mass against $\Omega h^2$ and $\langle\sigma v \rangle$. The cross section is rescaled with a factor $\xi^2=\left( \Omega h^2_{\textrm{Model}}/\Omega h^2_{\textrm{Planck}} \right) ^2$ when $\Omega h^2$ is smaller than the measured Planck value \cite{Aprile:2015uzo}. The left panel represents the 100\% flux scenario while the right panel represents the 40\% flux scenario.}
\end{figure}
\begin{figure}
\centering
\begin{subfigure}{.5\textwidth}
  \centering
  \includegraphics[width=\textwidth]{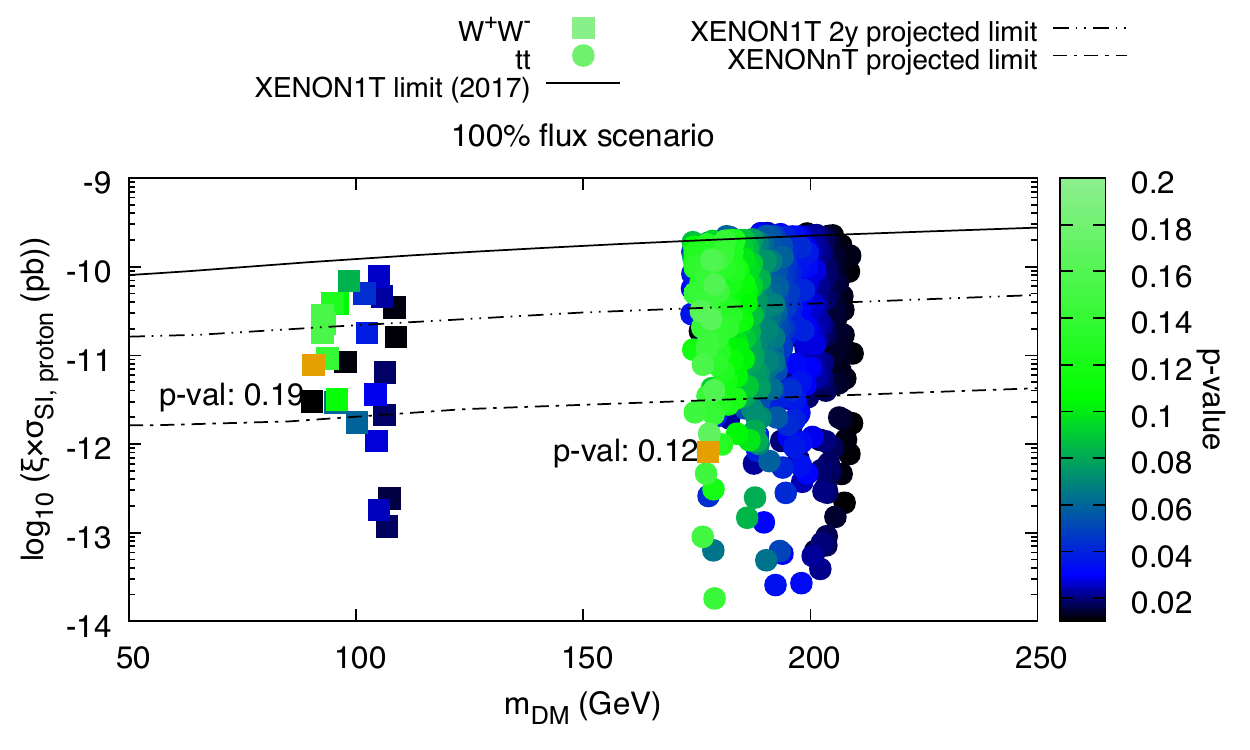}
  \caption{Neutralino mass against $\xi \sigma_{\textrm{SI,proton}}$\\for the 100\% flux scenario}
  \label{fig:neutralino_proton_sigma_si}
\end{subfigure}%
\begin{subfigure}{.5\textwidth}
  \centering
  \includegraphics[width=\textwidth]{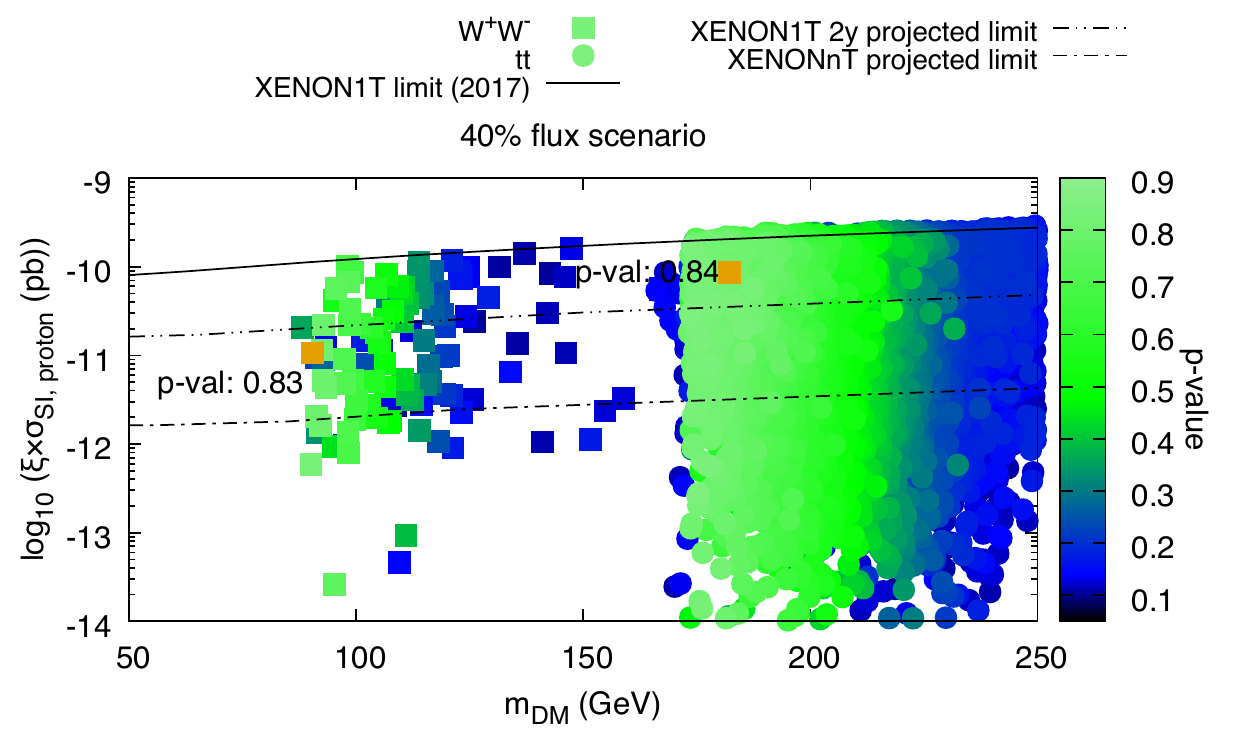}
  \caption{Neutralino mass against $\xi \sigma_{\textrm{SI,proton}}$  \\for the 40\% flux scenario}
  \label{fig:neutralino_proton_sigma_si_40p}
\end{subfigure}
\begin{subfigure}{.5\textwidth}
  \centering
  \includegraphics[width=\textwidth]{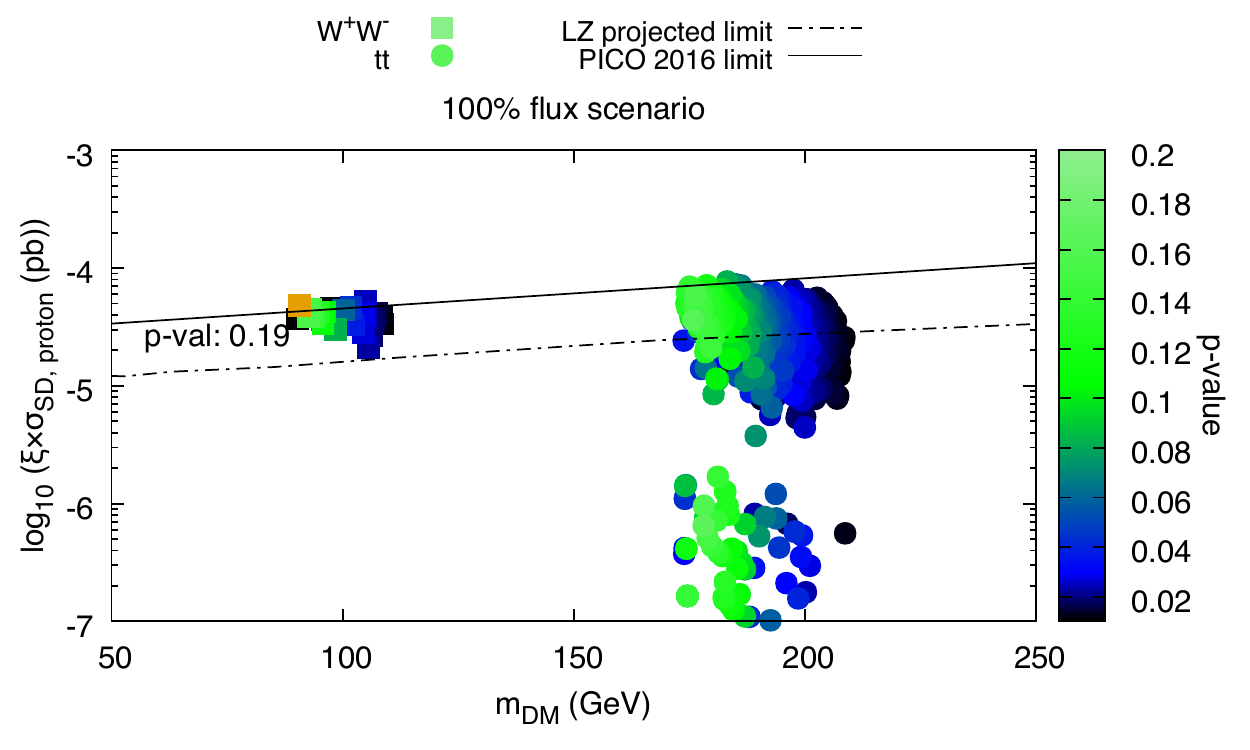}
  \caption{Neutralino mass against $\xi  \sigma_{\textrm{SD,proton}}$ \\for the 100\% flux scenario}
  \label{fig:neutralino_proton_sigma_sd}
\end{subfigure}%
\begin{subfigure}{.5\textwidth}
  \centering
  \includegraphics[width=\textwidth]{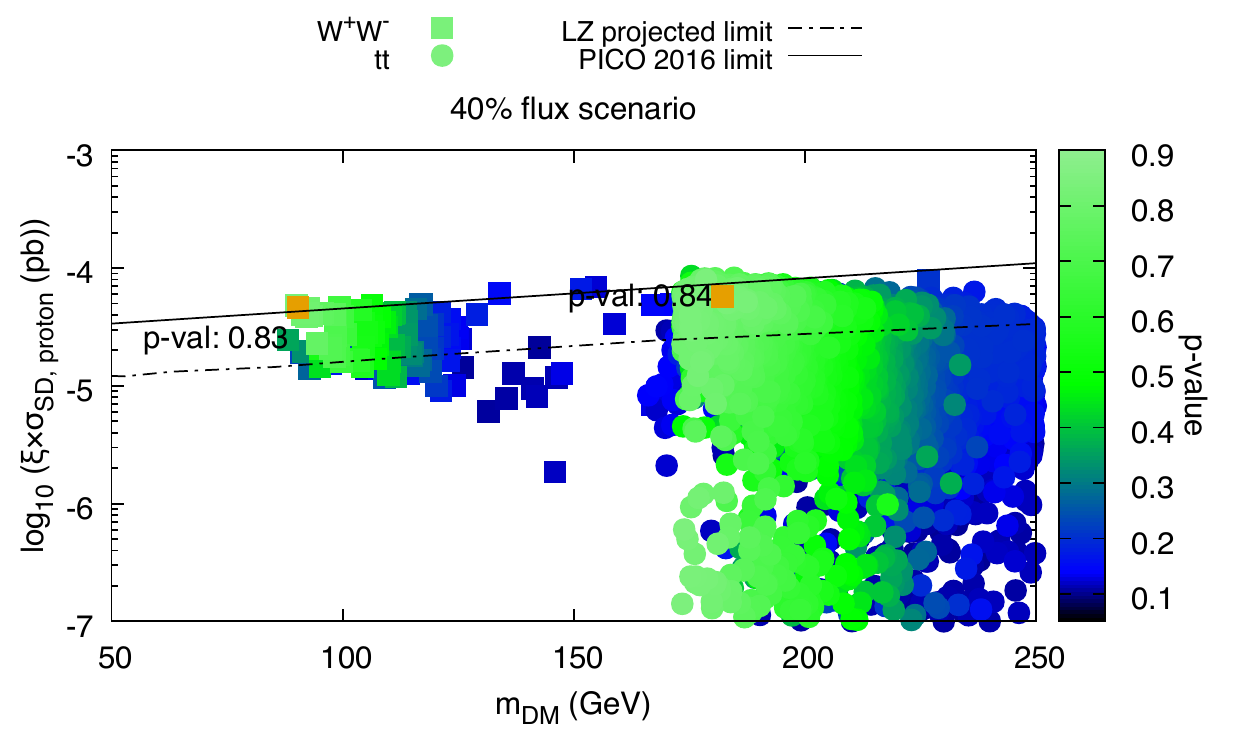}
  \caption{Neutralino mass against $\xi \sigma_{\textrm{SD,proton}}$ \\for the 40\% flux scenario}
  \label{fig:neutralino_proton_sigma_sd_40p}
\end{subfigure}
\caption{List of dark matter proton cross section plots. The cross sections are rescaled with a factor $\xi=\Omega h^2_{\textrm{Model}}/\Omega h^2_{\textrm{Planck}}$ when $\Omega h^2$ is smaller than the measured Planck value \cite{Aprile:2015uzo}. The left panel represents the 100\% flux scenario while the right panel represents the 40\% flux scenario. }
\label{fig:global_figures}
\end{figure}
Figures \ref{fig:neutralino_oh2} and  \ref{fig:neutralino_oh2_40p} show the relic density of the obtained models. We are agnostic about the cosmological model that gives rise to the DM abundance we observe today, therefore we do not imply any constraints on $\Omega h^2$. The relic density of models that have a high p-value is around the measured value. This is remarkable since the MSSM has many regions without a strong correlation between the relic density and the present-day DM annihilation cross section. For the same annihilation cross section, co-annihilation processes in the early Universe can result in vastly different relic densities. 
\newpage
The values for $\xi^2\langle\sigma v\rangle$ for the obtained models are shown in figures \ref{fig:neutralino_sv} and \ref{fig:neutralino_sv_40p}. In contrast to simplified model DM explanations of the GCE, the photon flux is  below the current {\it Fermi}-LAT limits from dwarf spheroidal galaxies \cite{Ahnen:2016qkx} and $\xi^2 \langle \sigma v \rangle$ is below the thermal annihilation rate of $3\times 10^{-26}$ cm$^{3}$s$^{-1}$. 

In figures \ref{fig:neutralino_proton_sigma_si} and \ref{fig:neutralino_proton_sigma_sd} the spin-independent (SI) and spin-dependent (SD) neutralino-proton cross-sections are shown. The projected and current sensitivities  (90\% CL exclusion limit) for the XENONnT, XENON1T, PICO and LZ experiments are also shown in these figures. We can observe that the complete $W^+W^-$ region for the $100\%$ flux scenario can be probed by the LZ experiment.

\clearpage
\begin{figure}[t]
	\centering
	\begin{subfigure}{.49\textwidth}
		\centering
		\includegraphics[width=\textwidth]{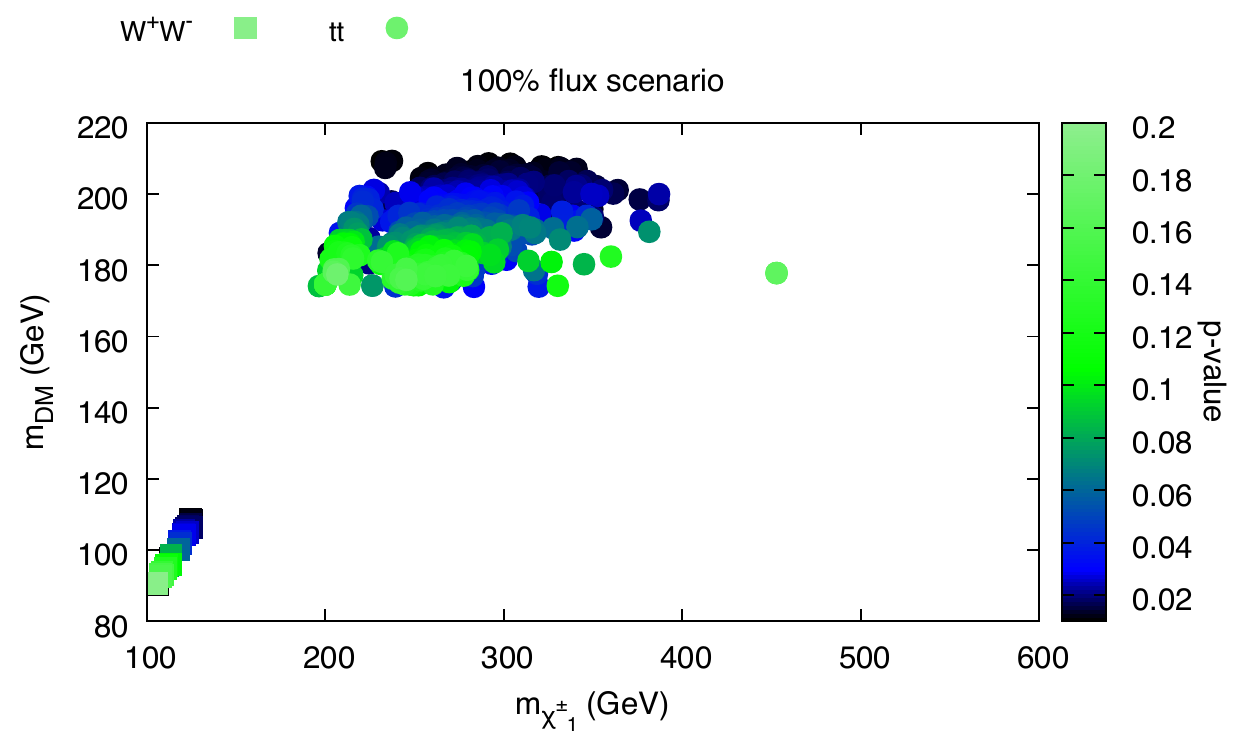}
		\label{fig:nmass-min-chargino}
	\end{subfigure}
    \begin{subfigure}{.49\textwidth}
		\centering
		\includegraphics[width=\textwidth]{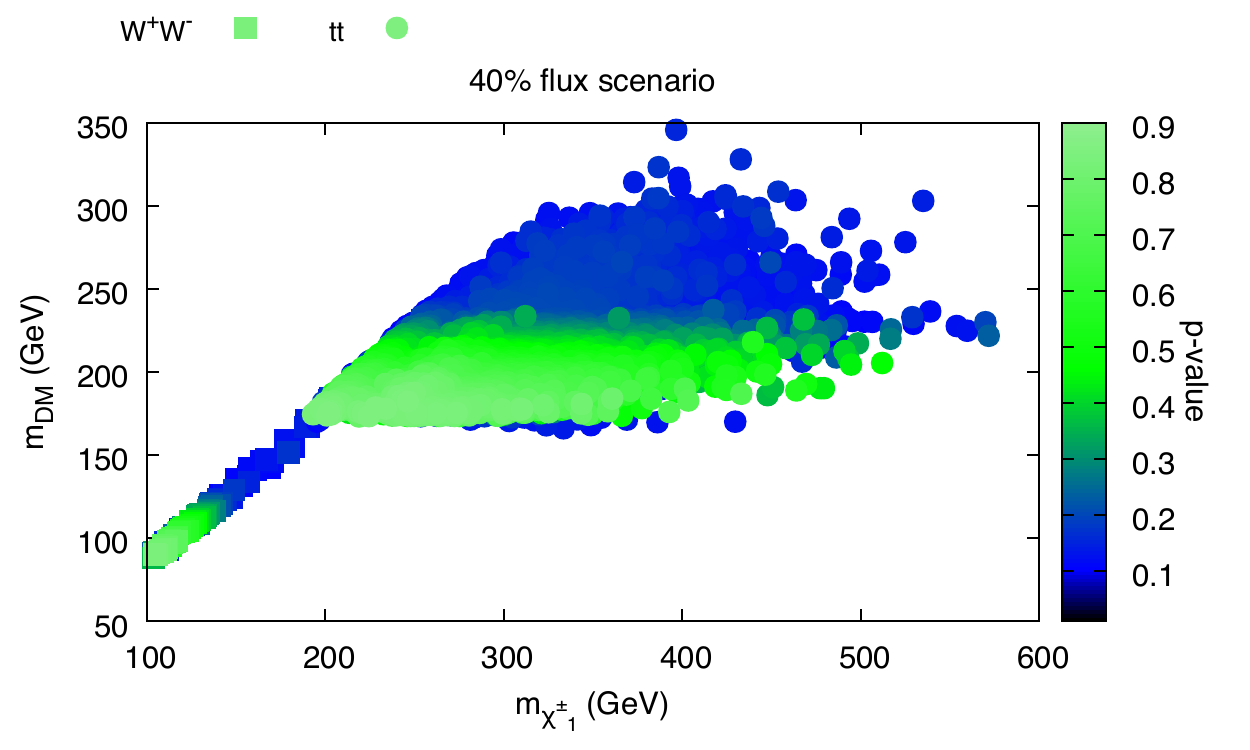}
		\label{fig:nmass-min-chargino_40p}
	\end{subfigure}
	\begin{subfigure}{.49\textwidth}
		\centering
		\includegraphics[width=\textwidth]{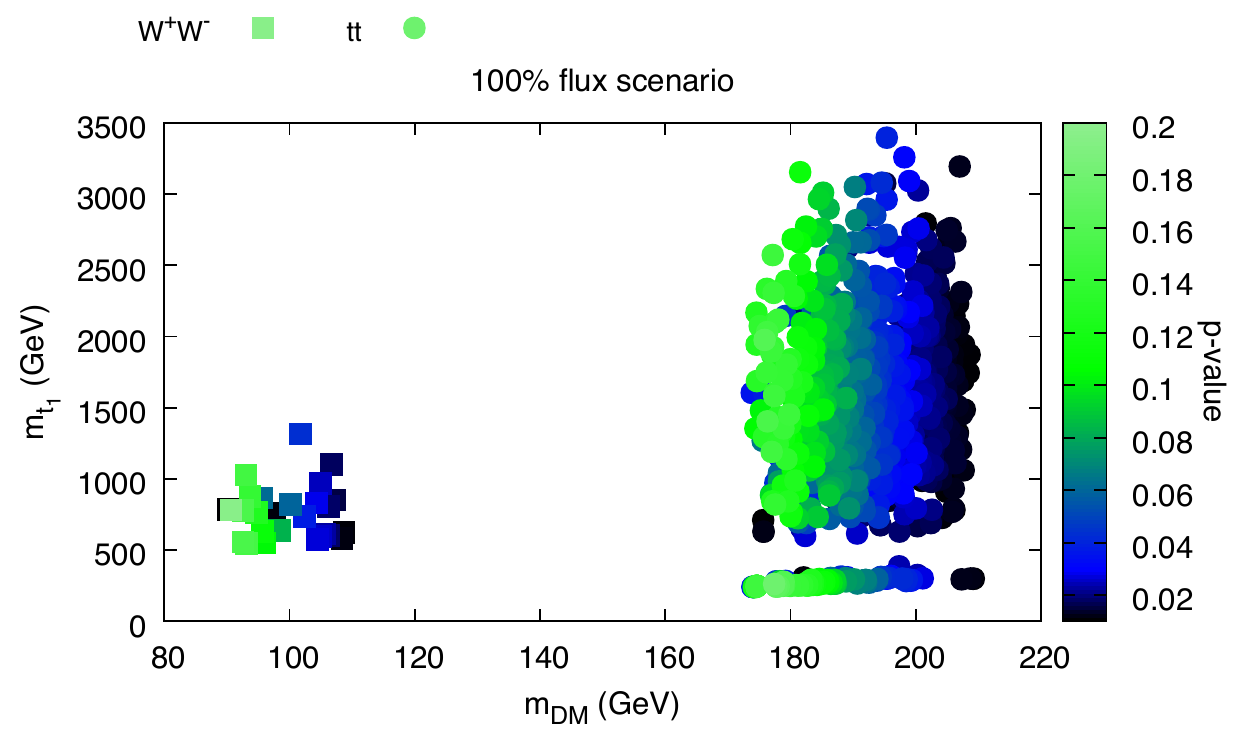}
		\label{fig:nmass-stop}
	\end{subfigure}
	\begin{subfigure}{.49\textwidth}
		\centering
		\includegraphics[width=\textwidth]{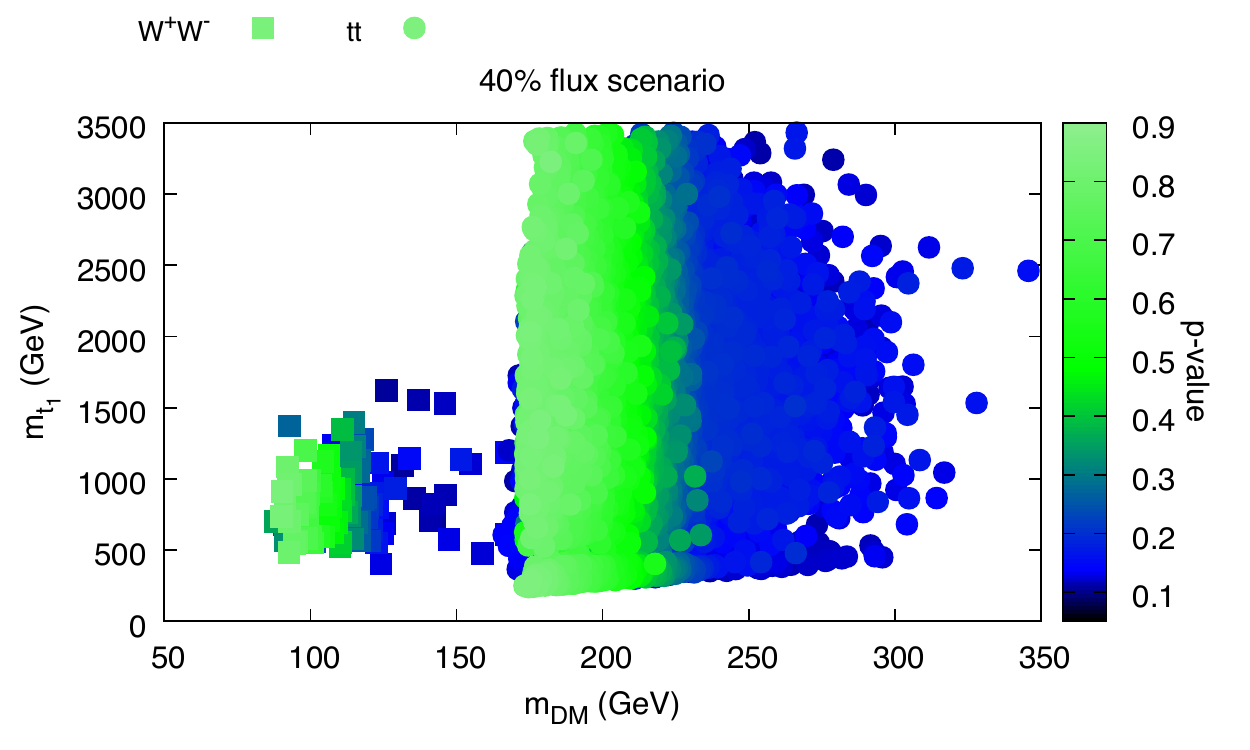}
		\label{fig:nmass-stop_40p}
	\end{subfigure}
\caption{Top lane: DM vs lightest chargino mass (in GeV) for the 100\% flux scenario (left) and the 40\% flux scenario (right). Bottom lane: DM vs lightest stop mass (in GeV) for the 100\% flux scenario (left) and the 40\% flux scenario (right). } 
    \label{fig:spectra_stop_chargino}
\end{figure}  

Finally, we turn our attention to LHC phenomenology. For the $W^+W^-$ region, the only relevant masses are given in the electroweak SUSY sector (the charginos and the neutralinos). In figure \ref{fig:spectra_stop_chargino} we show the mass of the co-annihilation partner (the chargino) vs the mass of the DM particle. For both the 100\% and the 40\% flux scenario, the lightest chargino $\tilde{\chi}^{\pm}_1$ and the next-to-lightest neutralino $\tilde{\chi}_2^0$ are close in mass with the DM particle. 
Their decays would create final state particles that are  relatively low energetic. 
In addition the production cross section of a wino-bino 
or higgsino-bino $\tilde{\chi}^0_2$  in processes like $\tilde{\chi}^0_2 \tilde{\chi}^{\pm}_1$ is reduced compared to the $100\%$ wino $\tilde{\chi}^0_2$ and $\tilde{\chi}^{\pm}_1$ production scenarios studied in most
LHC chargino-neutralino searches, such as in ref. \cite{CMS-PAS-SUS-16-048, ATLAS-CONF-2017-039}. To probe these models, one would need a dedicated compressed search as proposed in ref. \cite{vanBeekveld:2016hbo}. 
Another possibility is to search for signs of the heavier electroweak SUSY particles. For most models they are relatively light ($<500$ GeV) and decay to $\tilde{\chi}_1^+ W^-$ or $\tilde{\chi}_1^+ Z$. These could therefore trigger a signal in the dilepton and/or in the trilepton plus missing transverse energy channel, but have a reduced cross section as compared to the lighter mass states. 
\begin{figure}[t!]
	\centering
	\begin{subfigure}{.49\textwidth}
		\centering
		\includegraphics[width=\textwidth]{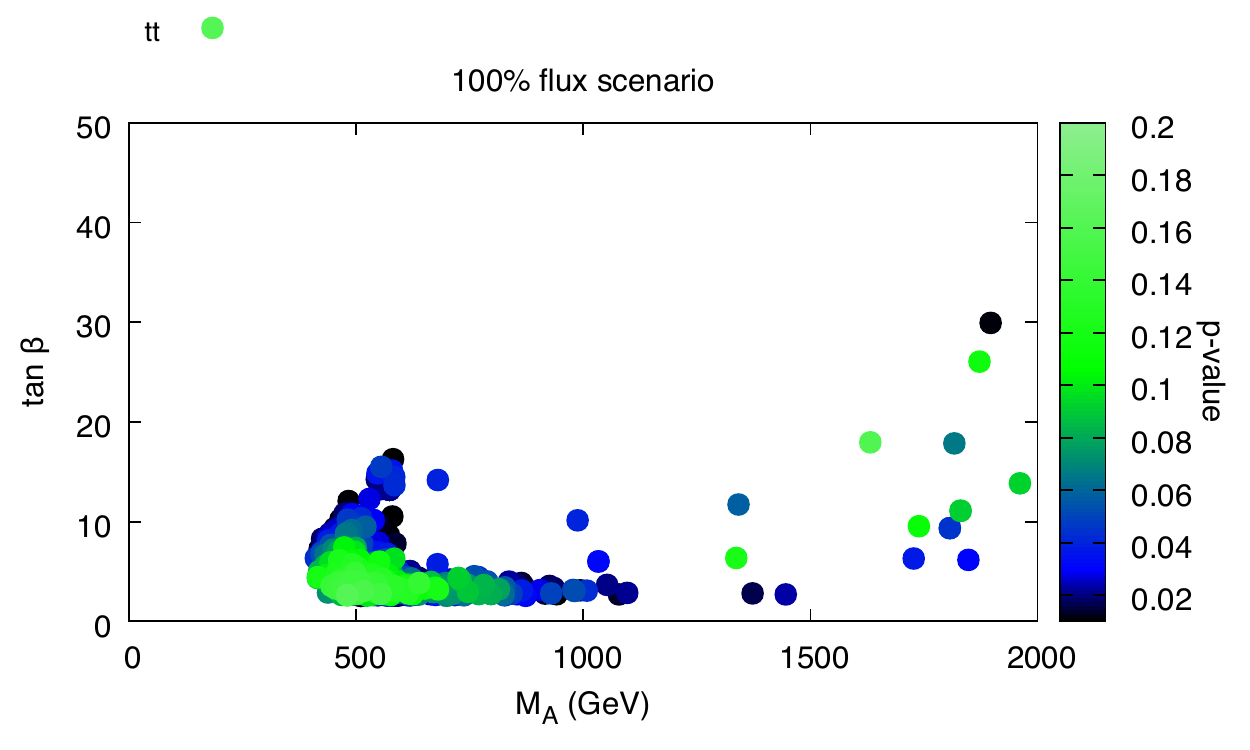}
	\end{subfigure}
    \begin{subfigure}{.49\textwidth}
		\centering
		\includegraphics[width=\textwidth]{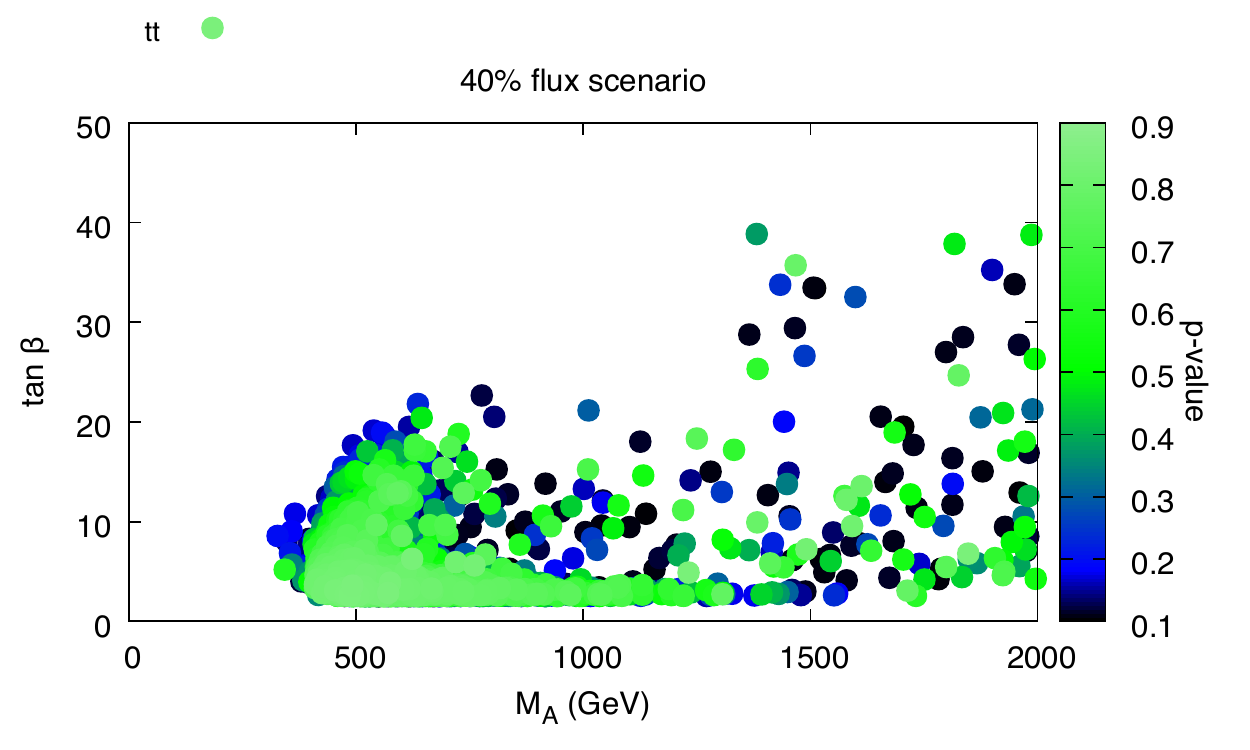}
	\end{subfigure}
	\begin{subfigure}{.49\textwidth}
		\centering
		\includegraphics[width=\textwidth]{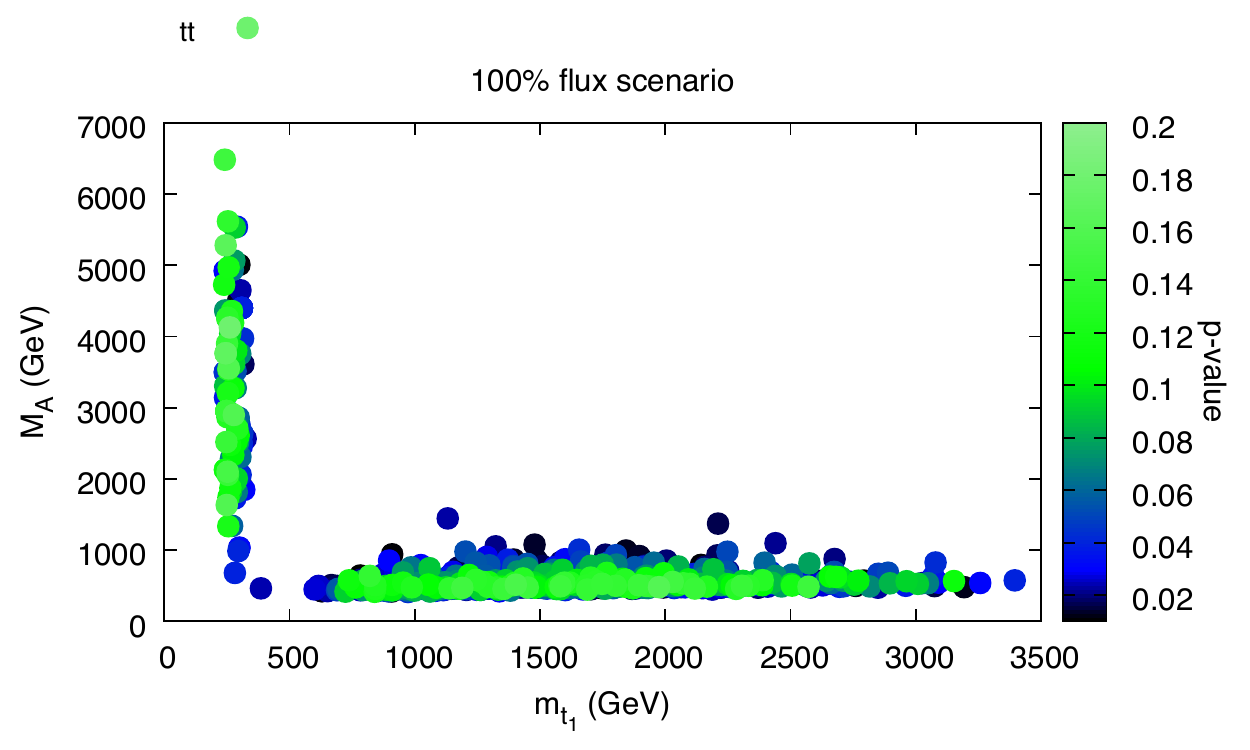}
	\end{subfigure}
	\begin{subfigure}{.49\textwidth}
		\centering
		\includegraphics[width=\textwidth]{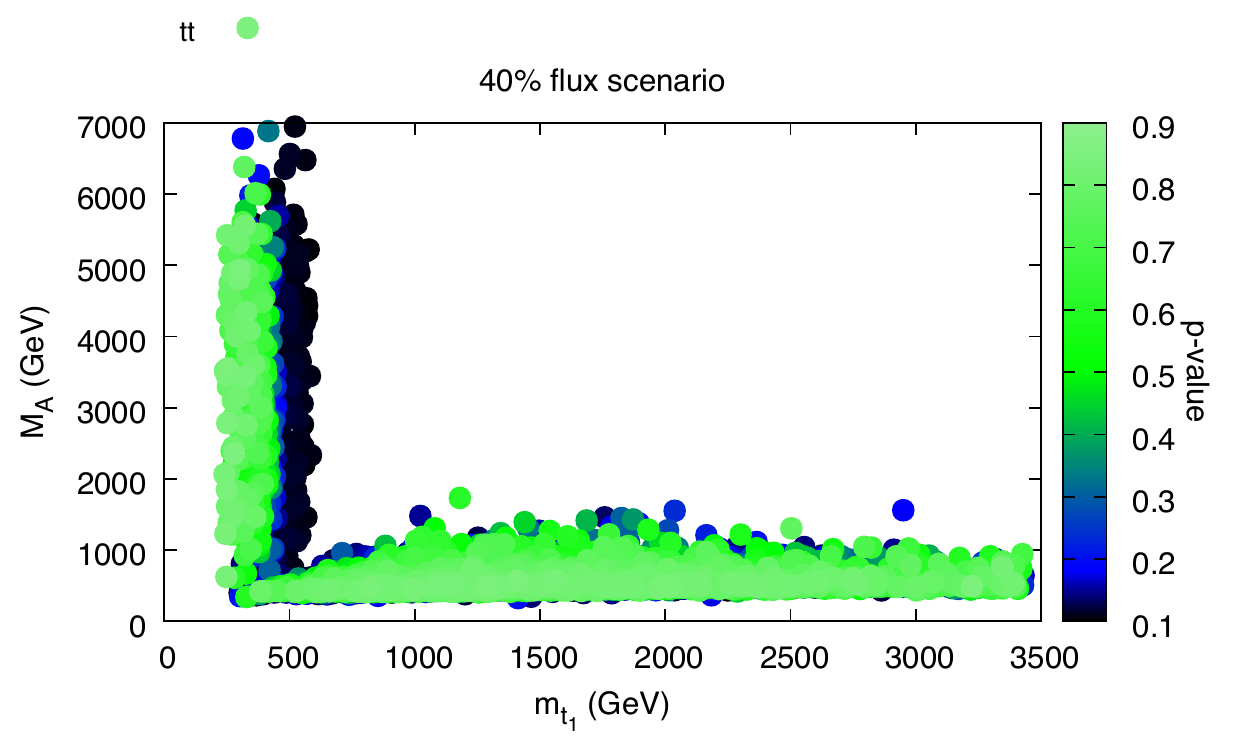}
		\label{fig:nmass-stop_40p}
	\end{subfigure}
    \begin{subfigure}{.49\textwidth}
		\centering
		\includegraphics[width=\textwidth]{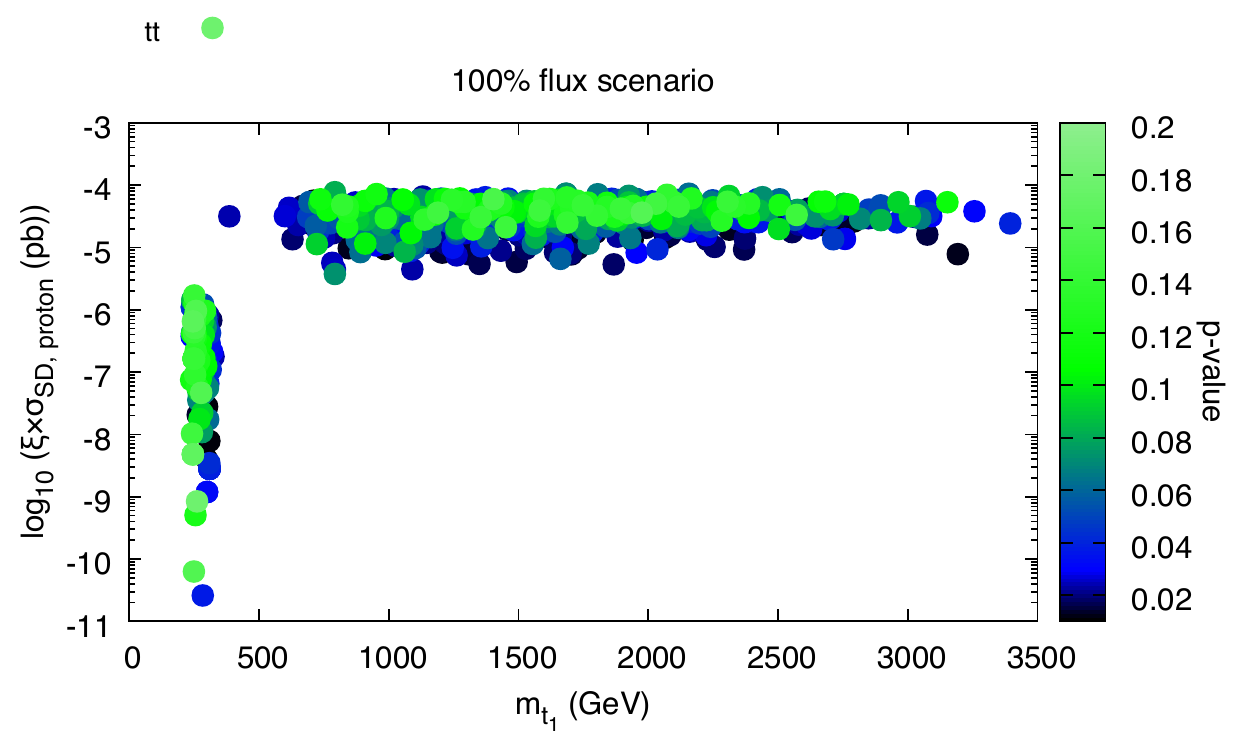}
	\end{subfigure}
    \begin{subfigure}{.49\textwidth}
		\centering
		\includegraphics[width=\textwidth]{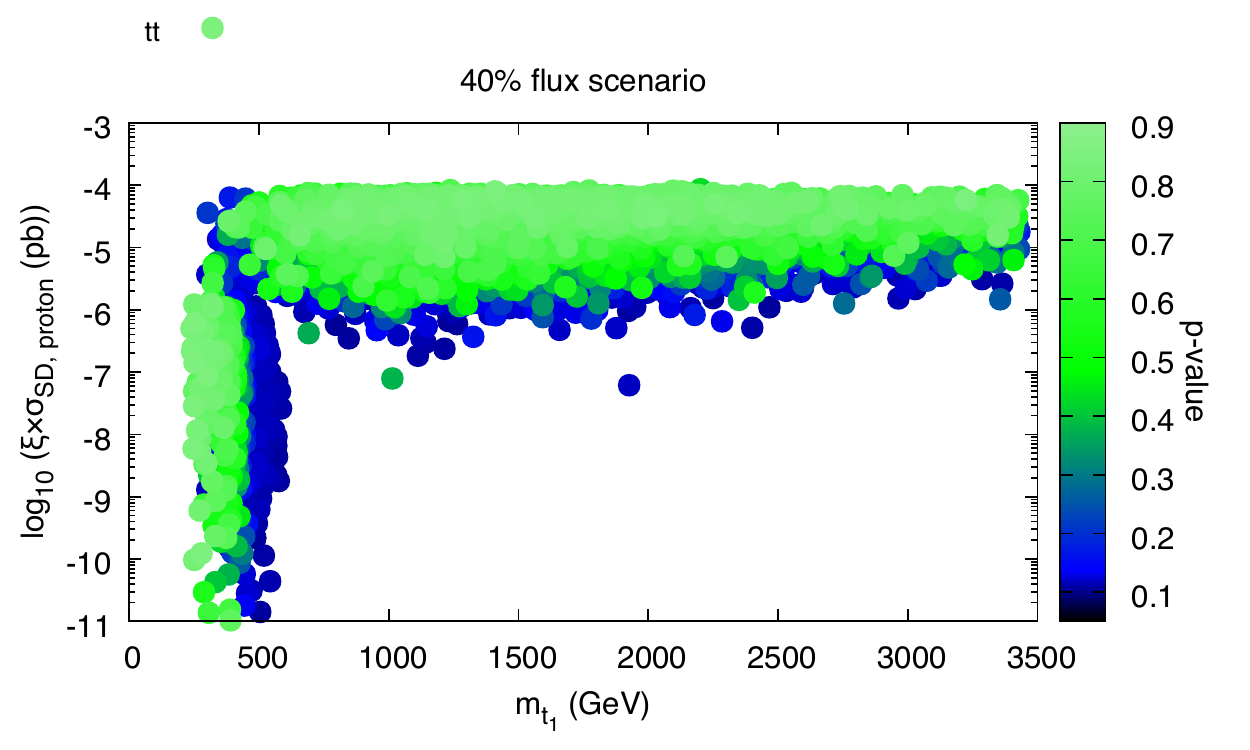}
	\end{subfigure}
\caption{Top lane: $M_A$ (in GeV) vs $\tan\beta$. Middle lane: Stop mass vs $M_A$ (both in GeV). Bottom lane: spin-dependent dark matter proton scattering cross section vs the stop mass (in GeV). The 100\% flux scenario is shown on the left side and the 40\% flux scenario is shown on the right side. Only the $t\bar{t}$ solutions are shown.} 
    \label{fig:ttbarma}
\end{figure}  

\textbf{In the $t\bar{t}$ region}, the DM particle annihilates predominantly to pairs of top quarks. There are three kinds of mechanisms responsible for the annihilation into top quark pairs. For DM particles that are almost purely bino, the annihilation happens exclusively via the hypercharge enhanced right-handed stop t-channel. These models all have a low stop mass (figure \ref{fig:ttbarma}, middle lane). A second region consists of DM particles that have a sizable higgsino component. These particles annihilate via the higgsino-pair annihilation channel via an s-channel $Z$-boson exchange\footnote{The annihilation to $t\bar{t}$-pairs is enhanced due to the helicity suppression that occurs for the annihilation to lighter fermions.}. These contributions become relevant when the lightest stop mass is $> 500$ GeV, since then the t-channel is then suppressed. If the neutralino mass is around 250 GeV a third possibility arises, which was first discovered in ref.  \cite{Freese:2015ysa}. For these neutralino masses, tops are produced via an s-channel exchange of the CP-odd Higgs boson with a mass around 500 GeV. The two higgsino enhanched regions are characterized by a low $\tan \beta<16$ and moderately low values for $M_A$. 
Regarding the relic abundance, we find points below the Planck bound as it can be seen in figure \ref{fig:neutralino_oh2} and \ref{fig:neutralino_oh2_40p} that correspond to solutions in which the lightest stop is almost degenerate with the DM particle and therefore they co-annihilate efficiently. It is remarkable that the point with the best p-value in this region has the correct relic abundance.
The $t\bar{t}$ models are not in any tension with direct detection experiments. The models that have a very low $\sigma_{{\rm SD},p}$ correspond to solutions where the neutralino is almost a pure bino and therefore the $Z$-boson coupling vanishes. This can also be seen in figure \ref{fig:ttbarma} (bottom lane), where $\sigma_{{\rm SD},p}$ is plotted against the stop mass. 

\begin{figure}[t]
	\centering
		\includegraphics[width=0.7\textwidth]{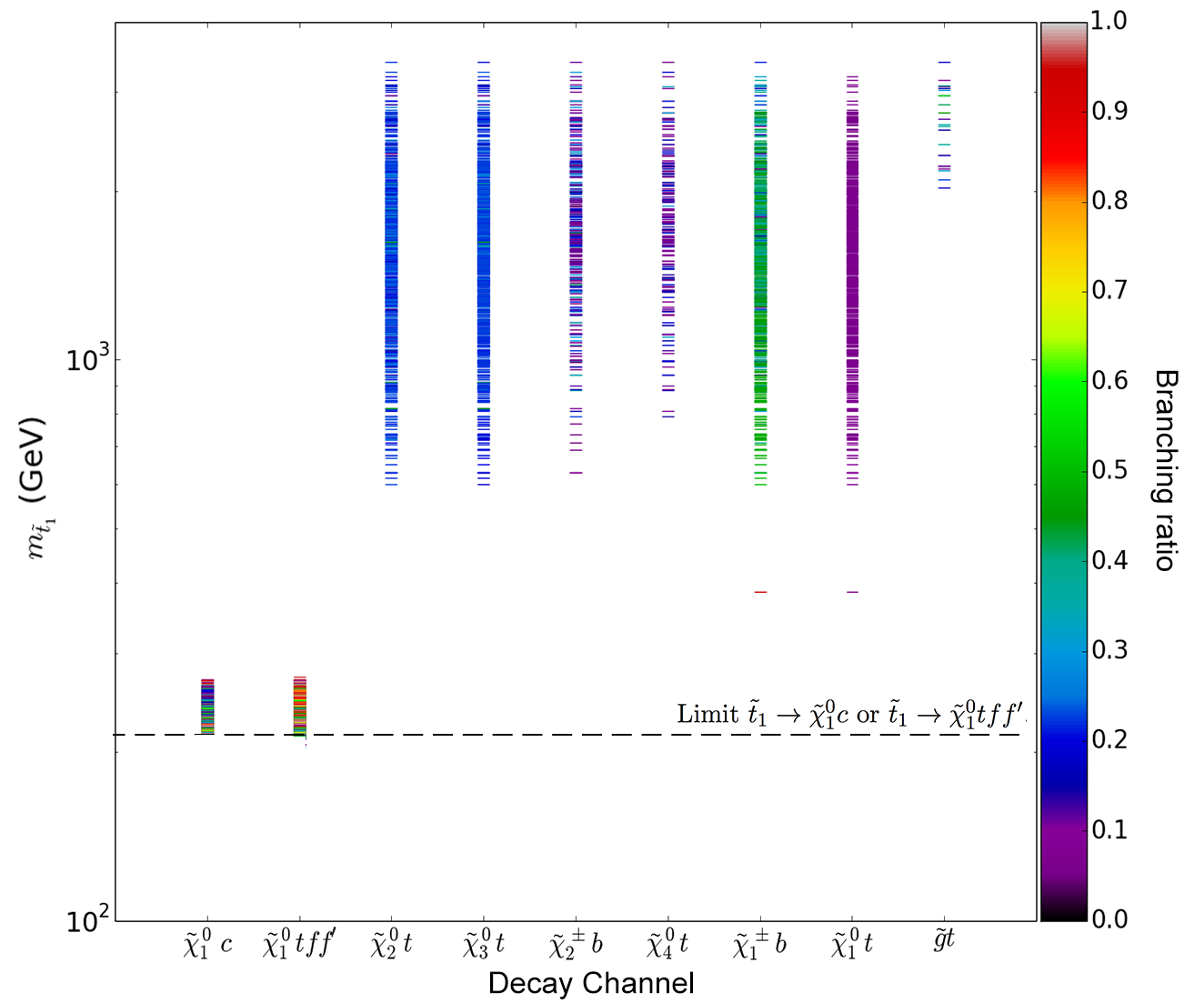}
	\caption{Decay spectrum for the $t\bar{t}$ solutions in the 100$\%$ flux scenario. The LHC limit for the decays $\tilde{t}_1 \rightarrow \tilde{\chi}^0_1 c$ and/or $\tilde{t}_1\rightarrow \tilde{\chi}_1^0 t f f'$ \cite{Aad:2014nra} is indicated by the dashed line. For heavier stops the simplified limits are not applicable.}
    \label{fig:stopdec}
\end{figure}

At the LHC the $t\bar{t}$ models can be probed via electroweak searches, searches for stop particles and searches for heavy Higgs bosons. Other SUSY particles are not relevant for the GCE interpretation. For some of the solutions, the lightest stop has a very low mass ($<500$ GeV, see figure \ref{fig:spectra_stop_chargino}), thus it would be produced copiously at the LHC. 
The decays for the $t\bar{t}$ models are shown in figure \ref{fig:stopdec}, together with the branching fractions as a color code. Models where the stop decays exclusively to the lightest chargino and a b-jet are excluded by the LHC bounds \cite{Sirunyan:2017wif}. If the chargino is heavier than the stop, the stop decays to three fermions and a neutralino or
to a charm and a neutralino. In these cases LHC bounds are
weaker and stops are only excluded if the stop mass is
below 220 GeV \cite{Aad:2014nra}. 
The obtained solutions are not in tension with the limits on heavy Higgs bosons \cite{Aaboud:2017hnm} or electroweak SUSY particles. 

Additionally the electroweak fine-tuning of the allowed models was calculated using the procedure from ref. \cite{vanBeekveld:2016hug,PhysRevLett.109.161802, PhysRevD.87.115028dd}. We find that the $W^+W^-$ region has a low fine-tuning ($<20$) overall, while the $t\bar{t}$ region has models which have a low fine-tuning as well. For our solutions, the higgsino component of the lightest neutralino essentially drives the value of the fine-tuning: the neutralinos with a higher higgsino component will result in models that have a lower fine-tuning. We therefore find that the $W^+W^-$ region has low fine-tuning. The $t\bar{t}$ models that have a low stop mass result in a higher fine-tuning, since the higgsino component of the neutralino is very small. The $t\bar{t}$ region does possess models with low values for the fine-tuning for the higgsino enhanced region. Details on the fine-tuning calculation can be found in ref. \cite{vanBeekveld:2016hug}. The fact that the models that can explain the GC excess also have a low fine-tuning is noteworthy.

\section{Conclusion}
\label{sec:conclusion}

In this analysis we verify that the phenomenological minimal supersymmetric standard model (pMSSM) can still explain the Galactic Center excess (GCE) using the Pass 8 data from {\it Fermi}-LAT and updated detector limits. We assume that dark matter annihilation is partially responsible for the excess flux between photon energies of 1 and 10 GeV. To account for the high energy tail of the excess, starting at photon energies of 10 GeV, a free power law was fitted. This resembles another astrophysical source to which we give no further interpretation. \\
Within the minimal supersymmetric extention of the Standard Model, two types of dark matter solutions are found that can explain the GCE:

\begin{itemize}
	\item \textbf{$W^+W^-$ region:} A neutralino with a mass between 80 and 120 GeV with the dominant neutralino annihilation channel being $W^+W^-$. The composition of the neutralino is mainly bino, with a smaller fraction of either wino or higgsino.
	\item \textbf{$t\bar{t}$ region:} A neutralino with a mass between 175 and 220 GeV annihilating predominantly to $t\bar{t}$. This solution splits up into two regions. If the neutralino is almost purely bino, the stop mass is typically close to the neutralino mass leading to a compressed stop-neutralino spectrum. If the neutralino has an enhanced higgsino component, the stop particle can have a mass up to a few TeV. The latter region is characterized by a heavy Higgs boson mass between 500 GeV and 1 TeV.  
\end{itemize}

It should be noted that in both regions, we find models that have a very low value for the electroweak fine-tuning. The direct dark matter detection experiments that provide limits on the spin-dependent dark matter-nucleus cross section will be able to probe the complete $W^+W^-$ region in the near future. The higgsino enhanced $t\bar{t}$ region can also be probed by these direct detection experiments. The LHC is able to probe the $t\bar{t}$ region and the $W^+W^-$ region via low mass stop searches, searches for heavy Higgs bosons and compressed chargino-neutralino searches.

\section*{Acknowledgements}
This paper uses the GCE extracted from \cite{1704.03910}, we thank Dmitry Malyshev for making the data available here https:/www-glast.stanford.edu/pub\_data/1220/. We thank Johannes Buchner for his advice on the statistical method to deal with GCE spectra. R. RdA, is supported by the Ram\'on y Cajal program of the Spanish
MICINN and also thanks the support of the Spanish MICINN's
Consolider-Ingenio 2010 Programme  under the grant MULTIDARK
CSD2209-00064, the Invisibles European ITN project
(FP7-PEOPLE-2011-ITN, PITN-GA-2011-289442-INVISIBLES, the  
``SOM Sabor y origen de la Materia" (FPA2011-29678), the
``Fenomenologia y Cosmologia de la Fisica mas alla del Modelo Estandar
e lmplicaciones Experimentales en la era del LHC" (FPA2010-17747) MEC
projects and the Spanish MINECO Centro de Excelencia Severo Ochoa del IFIC 
program under grant SEV-2014-0398. The work of GAGV was supported by Programa FONDECYT Postdoctorado under grant 3160153.
 SC and LH acknowledge the support within the "idark" program of the Netherlands eScience Center (NLeSC).
 M. van Beekveld acknowledges support by the Foundation for Fundamental Research of Matter (FOM), programme 156, "Higgs as Probe and Portal".
\bibliographystyle{ieeetr}
\bibliography{bibliography.bib}

\end{document}